\documentclass[12pt, referee]{aastex}
\usepackage[dvips]{color}

\shorttitle{Terahertz Water Masers}
\shortauthors{Neufeld et al.}

\begin{document}

\title{Terahertz Water Masers: II.  Further SOFIA/GREAT$^*$  Detections 
toward Circumstellar Outflows, and a Multitransition Analysis}
\author{David A.~Neufeld\altaffilmark{1}, Karl M.\ Menten\altaffilmark{2}, Carlos Dur\'an\altaffilmark{2,3},
Rolf~G\"usten\altaffilmark{2},  
Michael J.\ Kaufman\altaffilmark{4}, Alex~Kraus\altaffilmark{2}, Parichay Mazumdar\altaffilmark{2}, Gary J. Melnick\altaffilmark{5}, Gisela N.~Ortiz-Le\'on\altaffilmark{2},
Helmut~Wiesemeyer\altaffilmark{2}, and Friedrich Wyrowski\altaffilmark{2}}

\altaffiltext{*}{GREAT, the German REceiver for Astronomy at Terahertz frequencies, is a development by the MPI f\"ur Radioastronomie and the KOSMA/Universit\"at zu K\"oln, in cooperation with the DLR Institut f\"ur Optische Sensorsysteme.}
\altaffiltext{1}{Dept.\ of Physics \& Astronomy, Johns Hopkins University,
Baltimore, MD 21218, USA}
\altaffiltext{2}{Max-Planck-Institut f\"ur Radioastronomie, Auf dem H\"ugel 69, 53121 Bonn, Germany}
\altaffiltext{3}{European Southern Observatory, Alonso de Cordova 3107, Vitacura, Santiago, Chile}
\altaffiltext{4}{Dept.\ of Physics and Astronomy, San Jose State University, San Jose, CA 95192, USA}
\altaffiltext{5}{Center for Astrophysics $\vert$ Harvard \& Smithsonian, Cambridge, MA 02138, USA}

\begin{abstract}

Following up on our discovery of terahertz water masers, reported in 2017, we report
two further detections of water maser emission at frequencies above 1 THz.  Using the GREAT instrument on SOFIA, we have detected emission in the 1.296411 THz $8_{27}-7_{34}$ transition of water toward two additional oxygen-rich evolved stars, omicron Ceti (Mira) and R~Crateris, and obtained an upper limit on the 1.296~THz line emission from U Orionis.   Toward these three sources, and toward the red supergiant star VY Canis Majorae from which 1.296~THz line emission was reported previously, we have also observed several lower-frequency (sub)millimeter water maser transitions using the APEX 12~m telescope along with the 22 GHz transition using the Effelsberg 100-m telescope.  We have used a simple model to analyse the multi-transition data thereby obtained.  Adopting, as a prior, independent literature estimates of the mass-loss-rates in these four sources and in W~Hydrae, we infer water abundances in a remarkably narrow range: $n({\rm H_2O})/n({\rm H_2}) = 1.4 - 2.5 \times 10^{-4}$.  For $\it o$ Cet, VY CMa, and W Hya, the model is successful in predicting the maser line fluxes to within a typical factor $\sim 1.6 - 3$.  For R~Crt and U Ori, the model is less successful, with typical line flux predictions lying an order of magnitude above or below the observations; such discrepancies are perhaps unsurprising given the exponential nature of maser amplification.

\end{abstract}

\keywords{Masers --- ISM: molecules --- Submillimeter: ISM --- Molecular processes}

\section{Introduction}
  
Maser amplification is one of the more remarkable molecular phenomena observed in astrophysics. While population inversions have to be engineered by careful design in laboratory masers, they occur naturally for many molecular transitions at the low densities of the interstellar and circumstellar gas.  For dipole-allowed transitions of abundant molecules, maser amplification can lead to extraordinarily high brightness temperatures: in the case of the 22~GHz $6_{16}-5_{23}$ transition of water vapor, these may reach 10$^{14}$~K, enabling
observations to be performed by means of Very Long Baseline Interferometry (VLBI).  As discussed in Paper I (Neufeld et al.\ 2017), such observations yield a spectral resolving power, $\nu/\Delta \nu$, in excess of 10$^6$, and angular resolution at the submilliarcsecond level, providing a kinematic probe that is a powerful tool in a variety of astrophysical environments (e.g. Miyoshi et al. 1995; Reid et al. 2009).

Following the first detection of interstellar water vapor through its masing 22 GHz transition half a century ago (Cheung et al.\ 1969), roughly a dozen additional H$_2$O maser transitions have been discovered at ever higher frequencies.  These discoveries -- the details of which have been summarized in Paper I and will not be repeated here  -- have enabled investigations of the pumping mechanisms responsible for water maser emission in interstellar gas clouds, the circumnuclear disks of active galaxies, and circumstellar envelopes around evolved stars ({e.g.\ Humphreys et al. 2017; Baudry et al. 2018; Bergman \& Humphreys 2020}).  For most of the water maser transitions that have been detected in these environments, the observed emission can be accounted for by models that invoke collisional excitation as the dominant pumping mechanism (e.g.\ Neufeld \& Melnick 1991; hereafter NM91).

In Paper I, we presented the first detections of water maser transitions above 1 THz, which were obtained with the GREAT instrument on SOFIA.  In particular, the 1.296411 THz $8_{27}-7_{34}$ transition was detected towards three oxygen-rich evolved stars: W Hya, U Her, and VY CMa.  For the well-studied source W Hya, a simple model for the excitation of water vapor was able to account quantitatively for the observed fluxes in the 1.296 THz transition, the 22 GHz transition, and an even higher frequency transition ($8_{45} - 7_{52}$) of water at 1.884888 THz.  This model indicated that stimulated emission dominated the luminosity for the 1.296 THz transition, but did not contribute significantly to the 1.885 THz line luminosity (although the latter transition was predicted to be slightly inverted.)  A detection of maser emission in the 1.296 THz  transition has also been reported toward the star-forming region NGC 7538-IRS1 (Herpin et al.\ 2017).

In this paper, we report the detection of the 1.296 THz water line toward two additional evolved stars, $o$~Cet (Mira) and R~Crt -- and an upper limit on the line flux from U~Ori -- obtained using the GREAT spectrometer on SOFIA in December 2018.  In addition, we present the results of near-simultaneous observations of the 22~GHz transition at the Effelsberg 100 m radio telescope, and of 6 higher-frequency transitions at the APEX 12 m submillimeter telescope.  These ground-based ancillary data were obtained toward the three stars targeted by SOFIA in December 2018, plus VY CMa (which had been previously observed with SOFIA in 2017, yielding results discussed in Paper I).  The observations and data reduction procedures are described in Section 2, and results presented in Section 3.  In Section 4, we present a multitransition analysis of the water maser line fluxes measured in the new observations presented here and in previous studies.  A brief summary follows in Section 5.

\section{Observations and data reduction}
\subsection{SOFIA THz observations}
Table 1 lists the evolved stars toward which the water transitions were observed, along with the positions targeted, the variable star type, the spectral type, the period of the stellar variability, the source systemic velocity relative to the Local Standard of Rest (LSR), the estimated outflow velocity, the distance to the source, and the estimated mass-loss rate.  The source distances, $D$, in Table 1 were obtained from the revised parallaxes measured by {\it Hipparcos} (van Leeuwen 2007), those listed in Gaia DR2 (Gaia Collaboration 2018), and/or that determined by Zhang et al.\ (2012) with VLBI.  Where parallaxes from both satellite observatories were available (R~Crt and U~Ori), a weighted average was adopted (with the weighting inversely proportional to the square of the published uncertainties).  Estimates of the mass-loss rates for evolved stars have been obtained by a variety of different methods and often show a large dispersion for any given source.  Here, we have adopted estimates of the mass-loss rates for $o$ Cet\footnote{In the case of $o$ Cet, K98 identified two circumstellar outflows with different velocities and mass-loss rates: a slow wind with a terminal velocity of $2.4 \,\rm km\,s^{-1}$, and a fast wind with a terminal velocity of $6.7 \,\rm km\,s^{-1}$.  Planesas et al.\ (1990) had previously argued that the former was a standard spherical outflow while the latter was bipolar in nature.  While both components are clearly evident in the CO line profiles (K98, de Beck et al.\ 2010), the water maser lines appear only to trace the slow wind.  Even the 183 GHz H$_2$O emission, which --  in the other sources we observed -- spans almost the full range of velocities at which CO emission is detected, is only observed within the velocity range expected from the slow wind.}, 
U Ori, and R Crt from the study of Knapp et al.\ (1998; hereafter K98); these were all obtained using a consistent methodology based on measurements of CO line luminosities
.  However, where our estimate of the source distance differs from that assumed by K98, we scale the estimated mass-loss rate as $D^2$, because the CO line luminosities are expected to scale roughly linearly with the mass-loss rate.
Also shown are the date of each SOFIA observation, the corresponding stellar phase, the velocity of Earth's atmosphere relative to the LSR, the observatory altitude at the time of the observation, and the total integration time.  

The 1.296 THz water line was observed in the upper sideband of the 4G3 receiver of GREAT (Dur\'an et al.\ 2020) using the FFT4G backend; the latter provided 16,384 spectral channels 
with a spacing of 244~kHz.   At this frequency, the telescope beam has a diameter of $20^{\prime\prime}$ at half-maximum, as measured on the planet Mars.    The observations were performed in dual beam switch mode, with a chopper frequency of 1 Hz and the reference positions located 60$^{\prime\prime}$ on either side of the source along an east-west axis. 

The raw SOFIA data were calibrated to the $T_A^*$  (antenna temperature) scale, using an independent fit to the dry and the wet content of the atmospheric emission obtained by means of the task ``kalibrate" (Guan et al.\ 2012).  The inferred atmospheric transmission, which was dominated by the nearby telluric water line (Doppler-shifted relative to the source), varied from 0.28 to 0.82 with a median value of 0.42.
The forward efficiency was 0.97 and the main beam efficiency for the 4G3 band was 0.60, as calibrated on Mars.  The uncertainty in the flux calibration is estimated to be $\sim 20\%$ (Heyminck et al.\ 2012). 
Additional data reduction was performed using CLASS\footnote{Continuum and Line Analysis Single-dish Software.}.   This entailed the removal of a second-order baseline and the rebinning of the data to a channel width of 1.12~$\rm km \, s^{-1}$.

\subsection{APEX Submillimeter observations}
We used the APEX 12~m telescope to observe six additional transitions of water, three from ortho-H$_2$O and three from para-H$_2$O.  The complete set of observed transitions is listed in Table 2. 
The observations, of three stars targeted by SOFIA plus VY CMa, were conducted
between 2018 November 30 and December 6 under generally excellent weather conditions with the 12 m Atacama Pathfinder Experiment
submillimeter telescope (G{\" u}sten et al 2006). The atmospheric precipitable water vapor column (PWV) was between 0.2 and 0.5\,mm from November 30 to December 3 and between 0.6 and 0.8\,mm for the rest of the days.  The 183 GHz line was observed with the SEPIA180
receiver \citep{Belitsky2018}.  A highly evolved version of the first-light APEX submillimeter heterodyne instrument FLASH+ (Klein et al.\ 2014)
was used to observe the rest of the lines. It allows simultaneous observations of lines in the 345 and 460~GHz atmospheric windows. 

Calibration was obtained using the
chopper wheel technique under consideration of the very {significant variation of atmospheric opacity over the bandwidth, caused by telluric absorption in the very same H$_2$O lines that were our astronomical targets: the traditional chopper wheel technique widely used for the calibration of
(sub)millimeter single dish telescope observations \citep{Ulich1976}
compares the measurements over a receiver's full intermediate frequency (IF)
bandwidth (for both side bands) taken toward the blank sky, toward an ambient
load, and toward a cold load of either liquid N$_2$ or a cold head setup
(both at temperature ranges of 70--80 K). The sky temperature is fitted
by an atmospheric model and opacities are calculated to derive the correction
outside the Earth's atmosphere. In the past, this was often done with just one
opacity value for the full IF bandwidth of (nowadays) several GHz\footnote{With the significantly narrower bandwidths of past receivers, this was not an important issue in most cases.}. Clearly, this is
inadequate in cases for which the atmospheric transmission significantly
changes across a side band, which is always the case for H$_2$O lines.
Therefore the APEX data for all our observed lines have been calibrated by computing opacities per each
frequency channel\footnote{See the APEX Calibrator Manual, a permanent web document:
https://www.mpifr-bonn.mpg.de/technology/submm/calibrator\_manual\_latest}.
Signal and image sideband opacities are calculated separately and applied
according to the side band gain ratio.}

{For our observations of the 183.3101 GHz H$_2$O line, the SEPIA180 receiver was tuned so that the center of its upper (signal) side band corresponded to the rest frequency of this line with an adjustment 
for a source's LSR velocity. This places the SiO $v = 2, J-4-3$ line at GHz into the lower (image) side band. Since this maser line may have high intensities, its emission can potentially contaminate the other side band, despite the image band suppression, which is a nominal 18.5 dB on average 
\citep{Immer2016, Belitsky2018}. To examine whether this SiO line may have caused potential problems with our H$_2$O
line's signal band spectrum, we examined the image sideband data for all our sources to determine the SiO line's breadth.}

{The SiO line is expected to appear at its rest frequency of 171.2752 GHz, 
corrected for a source's LSR velocity, in the image side band, which was centered at a 12 GHz lower frequency than the signal side band. This causes a strongly attenuated contribution of the velocity-inverted SiO line spectrum to appear in the 183 GHz H$_2$O line spectrum in the signal band with  velocities shifted to $59.4$~km~s$^{-1}$ lower values relative to the systemic velocity, effectively resulting in no contamination. Indeed, if one very closely examines the 183 GHz H$_2$O spectrum of VY CMa (Fig. 2), our source with by far the widest line breadth, one can notice the very faint contribution of the SiO line at velocities $< 15$~km~s$^{-1}$, lower than any H$_2$O emission.}

The radiation was
analyzed with eXtended bandwidth Fast Fourier
Transform Spectrometers (XFFTS), which provide 65536 frequency channels over
2.5~GHz (Klein et al. 2012).  To increase
the signal to noise ratio, the spectra were smoothed to effective
velocity resolutions appropriate for the measured linewidths,
typically $\approx 0.5$~km~s$^{-1}$. 
Pointing checks were carried out with the receivers tuned to the HCN $J=2-1$ line or the SiO $J=4-3$ line for SEPIA180 and the CO $J=3-2$ line for FLASH+, either on the targets themselves or on nearby sources (HCN) with strong emission in these lines.
The pointing was found to be
accurate to within a few arcsec, acceptable given the FWHM beam sizes, which
are $34^{\prime \prime} $ at 183 GHz and $13^{\prime \prime} $ at 475 GHz.  In Table 3 we present
our line intensities in a flux density scale (i.e., in Jansky
units). For the conversion from antenna temperature above the atmosphere,
$T_{\rm A}^*$, to Jy, we used aperture efficiencies measured in December 2018 on Mars, converted to Jy per K factors as collected on the APEX website\footnote{http://www.apex-telescope.org/telescope/efficiency}. From the scatter of individual measurements we estimate a calibration uncertainty of 20\%. The conversion factors are given in Table 2.

 
%
%
%

\subsection{Effelsberg radio observations}
We used the Effelsberg 100-m telescope to
observe the 22.23508 GHz $6_{16} - 5_{23}$ transition toward the three sources targeted by SOFIA plus VY CMa. 
Like the $8_{27}-7_{34}$ transition observed with SOFIA/GREAT, the $6_{16} - 5_{23}$ transition is from the ortho-water spin isomer.
Several observations of each source were performed between 2018 December 7 and 11 with the 22 GHz 
secondary focus high electron mobility transistor receiver, which gave a FWHM beam 
size of about $40''$. The receiver was connected to a FFT spectrometer 
providing a frequency resolution of 1.5 kHz, corresponding to 0.02 km~s$^{-1}$. 
Weather conditions during the observations were unstable and changed rapidly,
impairing the data quality and stability. The data were corrected for (varying) atmospheric 
opacity and variations of the telescope’s gain curve with elevation. For the conversion 
of the observed spectra into Jy units, suitable calibration sources like 3C 286 and NGC 7027
were observed to determine the telescope’s sensitivity (which was about 1 K/Jy for all 
observations).
Due to the unstable weather conditions, the uncertainty of the spectra is of the order
of 10--15\%, in some cases even worse.

\begin{figure}
\includegraphics[width=13 cm]{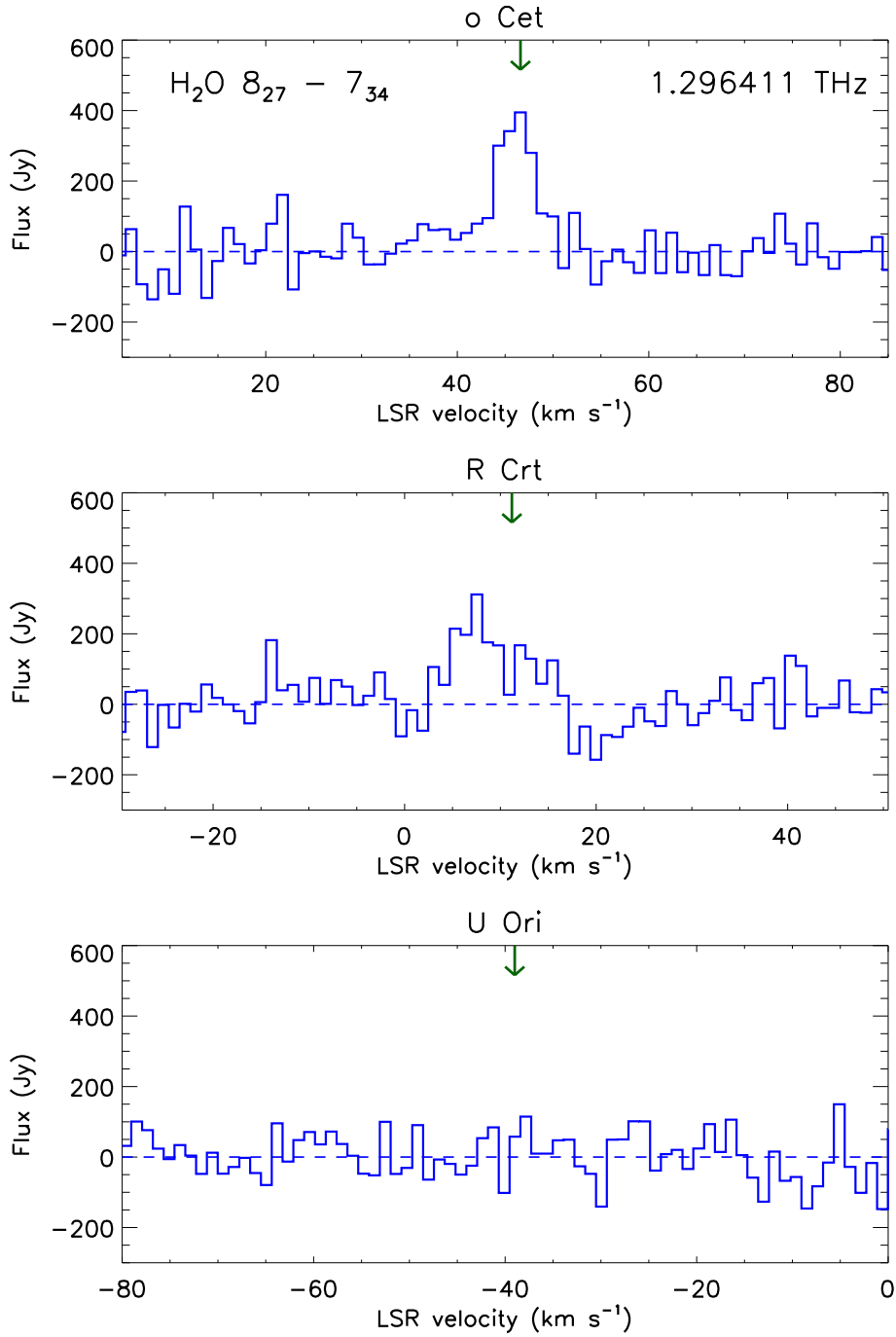}
\caption{Spectra covering the 1.296411 THz $8_{27}-7_{34}$ transition of water, observed by SOFIA/GREAT toward $o$~Cet, R~Crt and U~Ori.  Green arrows indicate the systemic velocity of each star.  Dashed lines indicate the baseline for each transition.}
\end{figure}

\begin{figure}

\includegraphics[width=13cm]{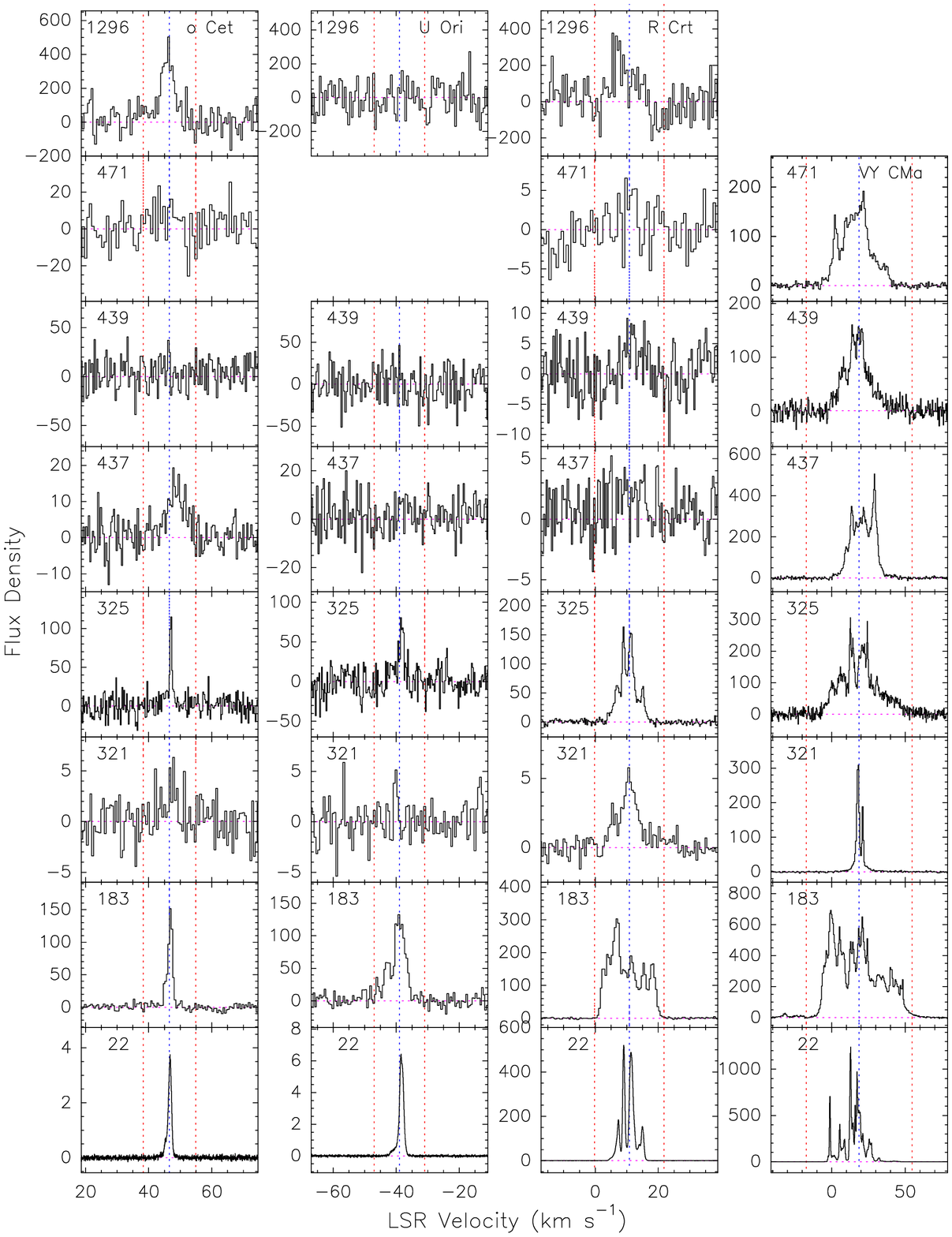}
\caption{Montage of spectra observed toward $o$~Cet, R~Crt and U~Ori, and VY CMa, combining the results of observations with SOFIA/GREAT, the APEX 12 m and the  Effelsberg 100 m telescopes. H$_2$O lines are identified by their frequencies (indicated in the upper left corner of each panel, rounded to the nearest GHz).
For each star, the vertical dotted
line marks the systemic LSR velocity, while the left and right vertical dotted lines mark the {velocity range
of the observed CO emission} (see Table 1). The horizontal dotted lines indicate the baseline level.}

\end{figure}

\section{Results}

In Figure 1, we present the reduced 1.296 THz water spectra obtained by SOFIA/GREAT, providing clear detections of the 1.296 THz water line toward $o$~Cet and R~Crt.  In Figure 2 we show the SOFIA/GREAT spectra along with those obtained for other water maser lines using Effelsberg and APEX.   In Table 3, we present the line-integrated fluxes measured for each source, the velocity range over which emission was detected, the peak flux, $S_p$, the LSR velocity where the peak flux is observed, $v_p$, and the implied isotropic luminosity for each transition, $L_p$ (in photons s$^{-1}$), given the source distances in Table 1.

\section{Discussion}

\subsection{Multitransition model}

In Paper I, we presented a simple model for the emission of maser radiation by  
one of the sources that we observed, W~Hya.   We adopted the gas temperature and velocity profiles given by Khouri et al.\ (2014), and used an excitation model to compute the radial dependence of several quantities for each transition: the 
maser optical depth along a tangential ray in the unsaturated limit, $\tau({\rm unsat})$; the rate of spontaneous radiative decay per unit volume $\Phi_p({\rm spont})$; the photon emission rate per unit volume in the limit of saturation, $\Phi_p({\rm sat})$; and the actual photon emission rate, $\Phi_p$.   
The quantities $\tau({\rm unsat})$ and $\Phi_p({\rm sat})$ were computed using the method discussed in NM91, and $\Phi_p$ was estimated by considering maser amplification along tangential rays.  As discussed in Paper I, our simple model neglects the effects of radiative excitation or pumping through vibrational states.   Except for one transition (the 437 GHz water line), all the water lines we observed are categorized as collisionally-pumped masers (Gray et al.\ 2016); the effects of radiative excitation are therefore expected to be modest.  {Previous studies (e.g. Yates et al.\ 1997,  Bergman \& Humphreys 2020)  have shown that the effect of dust continuum radiation is generally to reduce  the luminosity of collisionally-pumping water maser lines. While this suppression may have significant effects on the luminosity of high-gain, unsaturated masers (e.g.\ Bergman \& Humphreys 2020, their Figure 13), only modest reductions are expected for saturated masers unless the mean intensity of the continuum radiation exceeds $\sim 10\%$ of the Planck function at the gas kinetic temperature (e.g.\ NM91, their Figure 7); this condition is not expected to be met at far-infrared/submillimeter wavelngths in dusty circumstellar outflows.\footnote{Naturally, if the intensity of the continuum radiation is {\it equal} the Planck function at the gas kinetic temperature, then the level populations are driven to LTE and all maser action inevitably ceases.}}

In the present study, we have now included the para-H$_2$O spin isomer with an assumed ortho-to-para ratio of 3, and have expanded the model predictions to include all ten of the maser lines listed in Table 2.  {The model now includes 120 states each for both ortho- and para-H$_2$O.}  For each star we considered, we constructed a grid of 960 models with mass-loss rates, $\dot M$, ranging from $10^{-8.9}$ to $10^{-3}\, M_\odot \, \rm yr^{-1}$, and water abundance, $x_{\rm H2O} = n({\rm H_2O})/n({\rm H_2})$, ranging from $10^{-4.5}$ to $10^{-3}$.
The latter is assumed to be constant throughout the outflow; this assumption is consistent the fact that the photodissociation radius for water is expected to 
greatly exceed the radius of the maser emission region.
For both parameters, the grid points were spaced by 0.1 dex.  Separate grids were obtained for each source, with the appropriate temperature and velocity profiles specified in Table 4.  The set of sources consisted of the three stars observed by SOFIA in December 2018, plus W~Hya and VY~CMa.  Earlier SOFIA/GREAT observations of these last two stars were reported in Paper I.  W~Hya and VY~CMa are particularly valuable sources for testing the maser excitation model, because a large number of water maser transitions\footnote{With the new APEX observations reported here, all ten of transitions listed in Table 2 have now been detected toward VY CMa.} has been detected toward each of them (Menten et al.\ 2008; hereafter M08).  

\subsection{Best-fit water abundance at constant mass-loss-rate}

\begin{figure}
\includegraphics[width=8 cm]{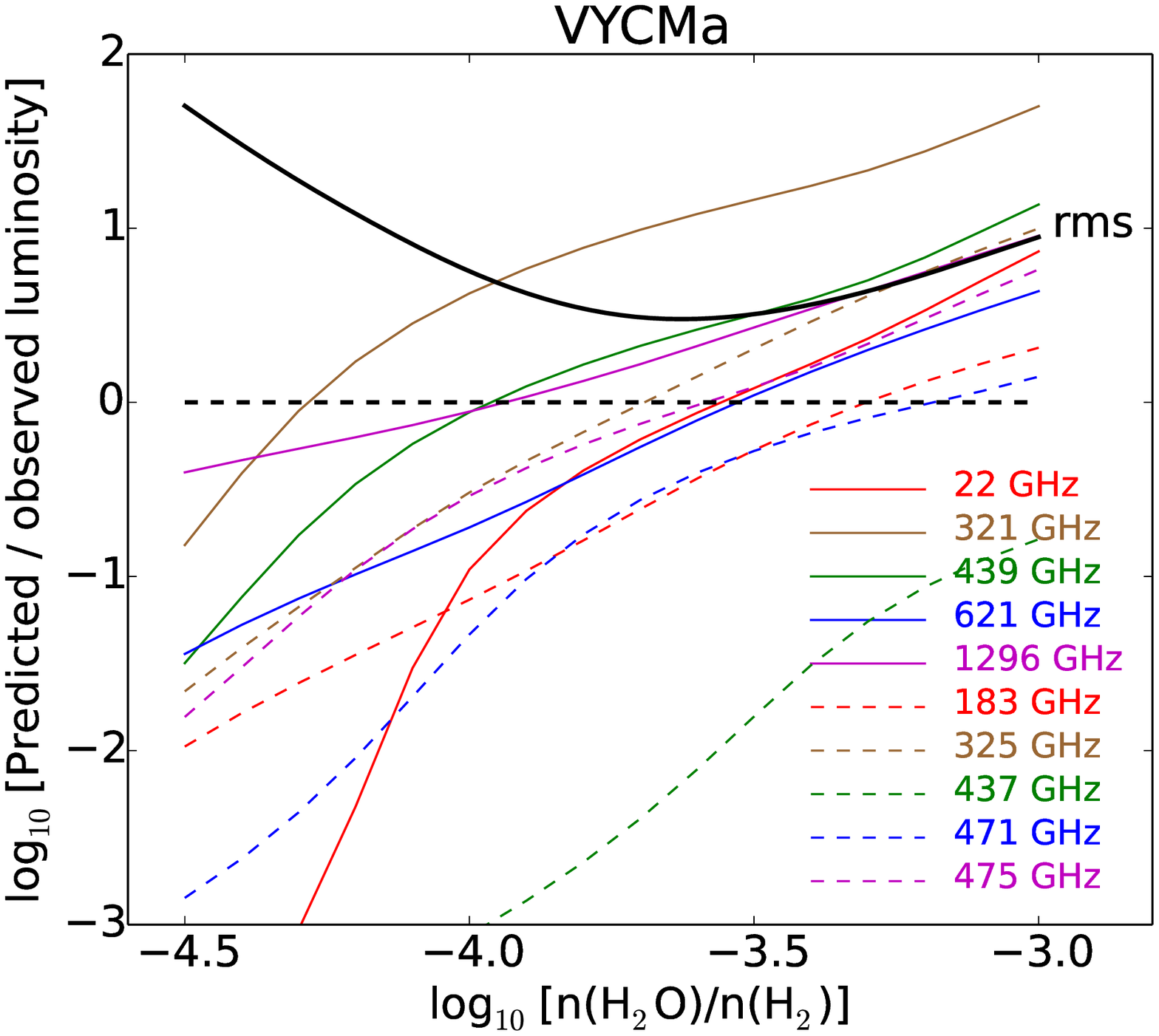}
\includegraphics[width=8 cm]{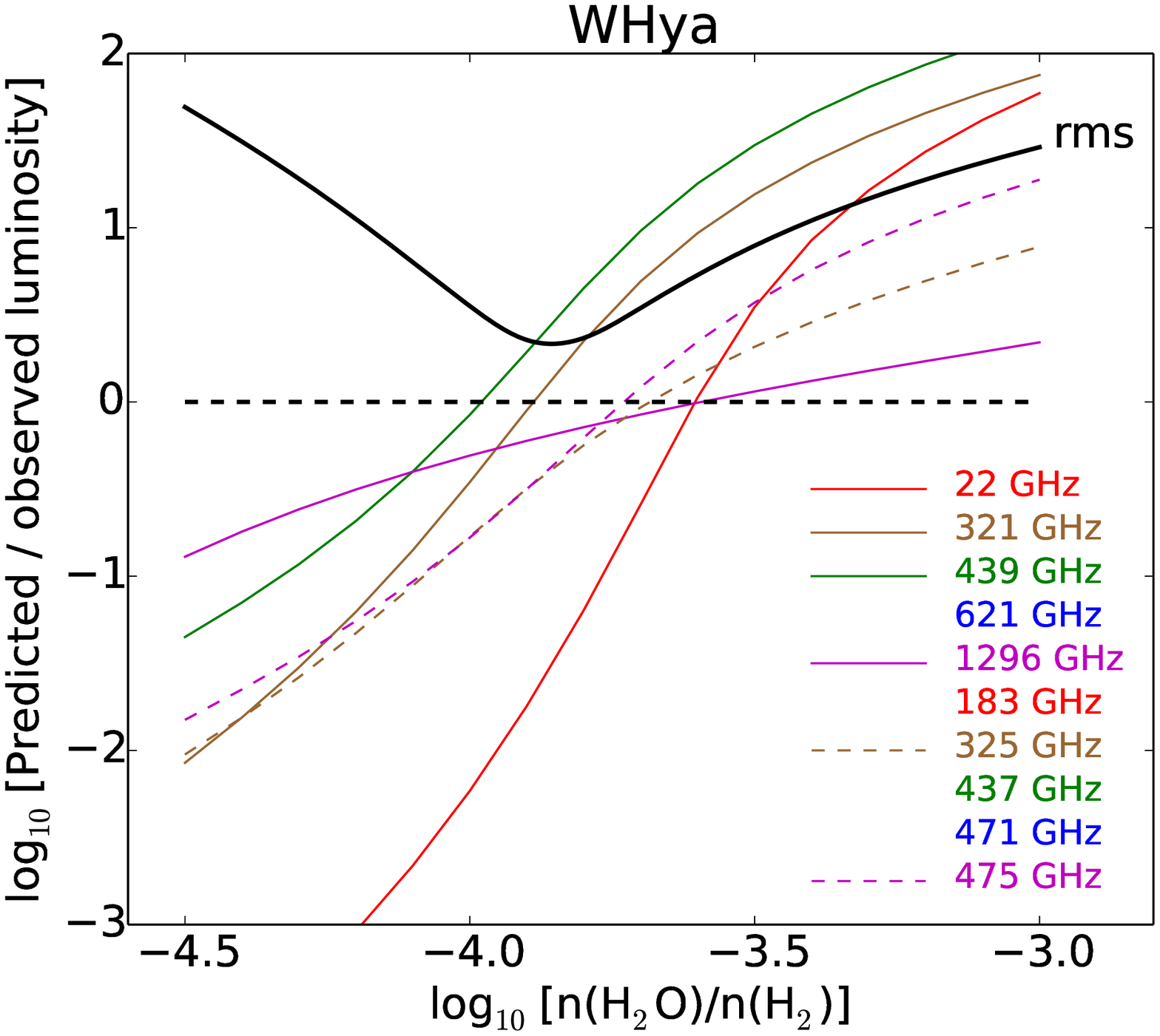}
\includegraphics[width=8 cm]{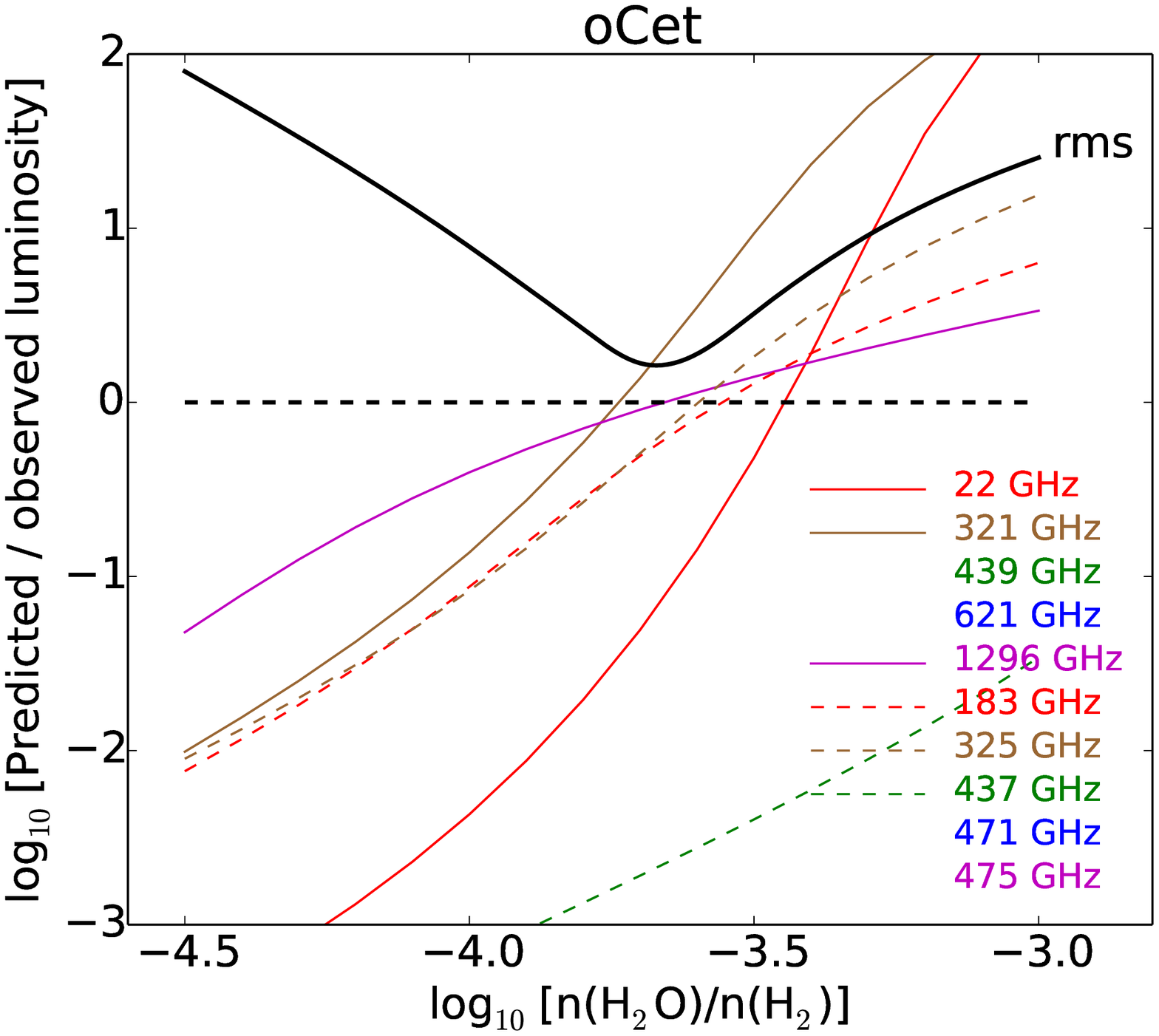}
\includegraphics[width=8 cm]{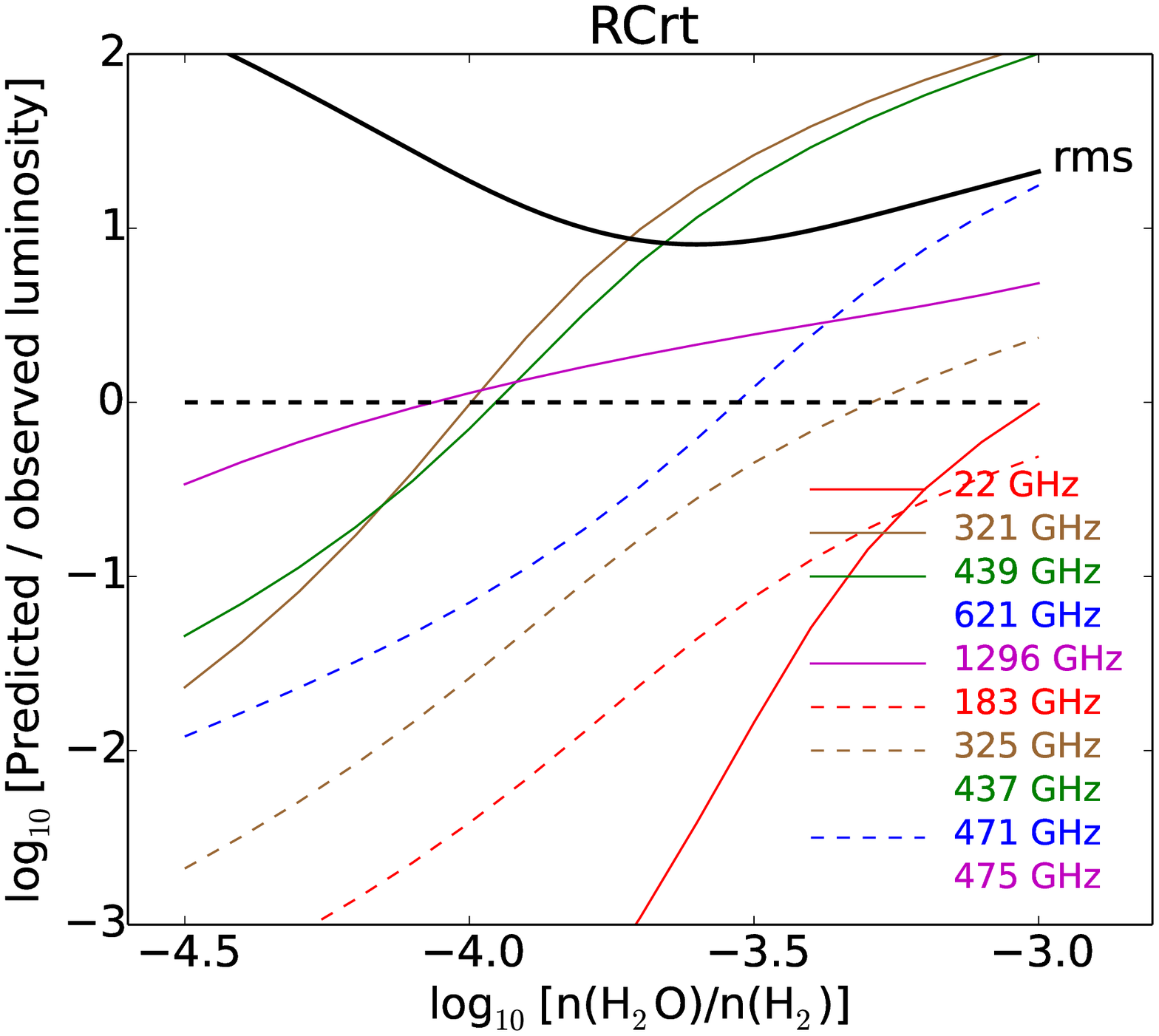}
\includegraphics[width=8 cm]{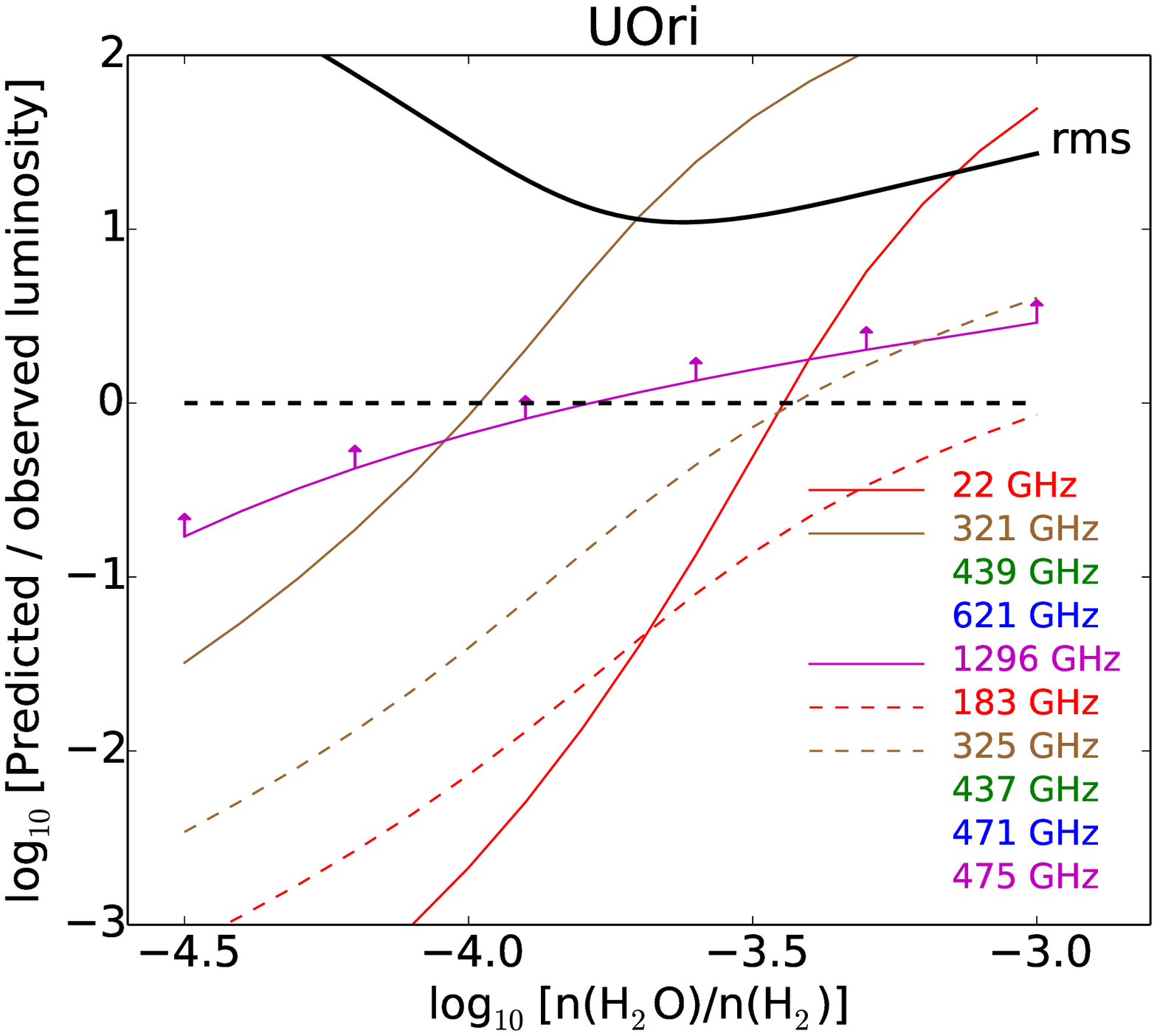}
\caption{Predicted-to-observed ratios for maser transitions observed toward
each source, as a function of the assumed water abundance $x_{\rm H2O} = n({\rm H_2O})/n({\rm H_2})$, given the mass-loss rates in Table 1.  Thick black curve: r.m.s value of ${\rm log}_{10}(F_{\rm obs}/F_{\rm mdl})$ (see text).}
\end{figure}

In confronting the model predictions with the data, we first adopted the mass-loss rates 
in Table 1 as a prior, and treated the water abundance as a free parameter.  Figure 3 shows
how the measured line fluxes, $F_{\rm obs}$, compare with the predictions of the model 
$F_{\rm mdl}$, given the distances tabulated in Table 1\footnote{In the case of W Hya (for which no new observations were obtained
in the present study), we adopted a distance of $138 \pm 12$~pc and a mass-loss rate of $2.45 \times 10^{-7}\,M_\odot\,\rm yr^{-1}$ (K98, corrected for the different distance estimate assumed here).  Based on our inclusion of the newer Gaia DR2 parallax for W~Hya, our adopted distance is significantly larger than the value of 78~pc assumed in Paper I.}.

Here, colored lines show the values of ${\rm log}_{10}(F_{\rm mdl}/F_{\rm obs})$ for the various maser transitions that have been detected. 
Solid lines apply to o-H$_2$O transitions, and dashed lines to p-H$_2$O transitions.  
For U~Ori, the curve shown for the 1296~GHz transition -- with upwards facing arrows -- corresponds the upper limit obtained on $F_{\rm obs}$ (Table 3); for $x_{\rm H2O} \ge -3.8$, the model prediction starts to exceed the observed upper limit.

Among the sources we consider here, VY CMa and W Hya have been the most extensively studied.  For both stars, however, the various transitions were not all observed at a single epoch, a significant shortcoming given the well-known variability of maser transitions.  In the case of VY CMa, seven of the ten maser transitions have been observed
in both the present study and by M08, with fluxes that varied by a factor of up to 1.4: for those transitions, we took the average of the photon luminosities
detected at those two epochs. 

There are two key features of the results shown here.  First, the 22~GHz transition (solid red line) typically shows a very strong dependence on the assumed water abundance.  This behavior is expected, because the 22~GHz transition exhibits the largest gain 
and typically has the largest ratio of $\Phi_p({\rm sat})/\Phi_p({\rm spont})$: the predicted
flux is therefore strongly dependent on our assumptions about the water abundance and other aspects of the model. Second, the 437~GHz line luminosity is severely underestimated for any reasonable water abundance; this behavior has been noted previously (Melnick et al.\ 1993, M08, Gray et al.\ 2016) and remains a puzzle.

As a measure of the goodness-of-fit for the model, we therefore computed the r.m.s value of
${\rm log}_{10}(F_{\rm mdl}/F_{\rm obs})$, averaging over all detected transitions but excluding the values obtained for the
22~GHz and 437~GHz transitions.  The solid black curve shows the result, ${\rm rms} \,({\rm log}_{10}[F_{\rm mdl}/F_{\rm obs}])$.   In Table 4, we list for each star the minimum values of 
${\rm rms}\,({\rm log}_{10}[F_{\rm mdl}/F_{\rm obs}])$ and the best-fit water abundances, $x_{\rm H2O}$, at which they are obtained.

\begin{figure}
\includegraphics[width=13 cm]{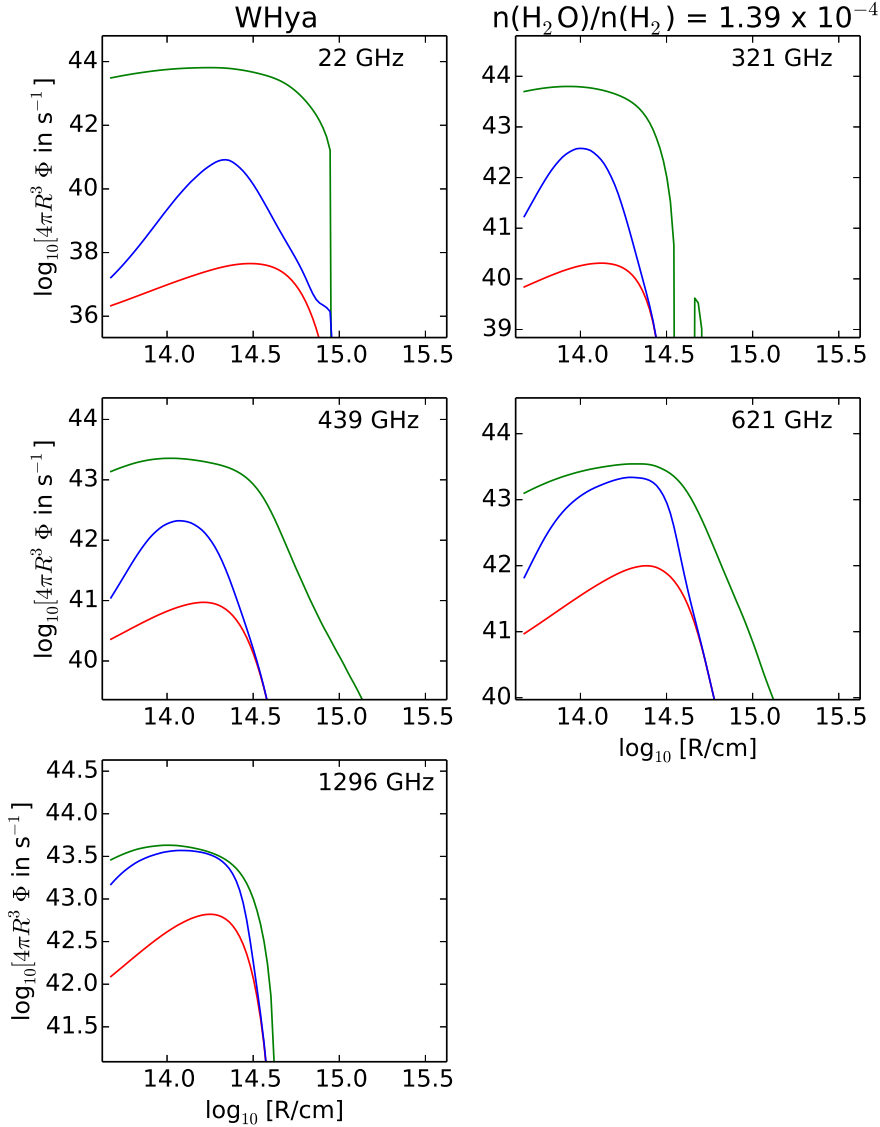}
\caption{Emissivity profiles for transitions of ortho-H$_2$O.  Here, we show how three predicted
quantities depend on the radial coordinate, $R$, for the best-fit model obtained for W Hya. 
Red curves: rate of spontaneous emission per logarithmic distance element, $4\pi R^3 \Phi_p({\rm spont})$, in units of photons s$^{-1}$.  Green curves: rate of photon emission per logarithmic distance element for saturated maser emission, $4\pi R^3 \Phi_p({\rm sat})$.  Blue curves: estimate of the actual photon emission rate per logarithmic distance element, $4\pi R^3 \Phi_p$.}
\end{figure}

\begin{figure}
\includegraphics[width=15 cm]{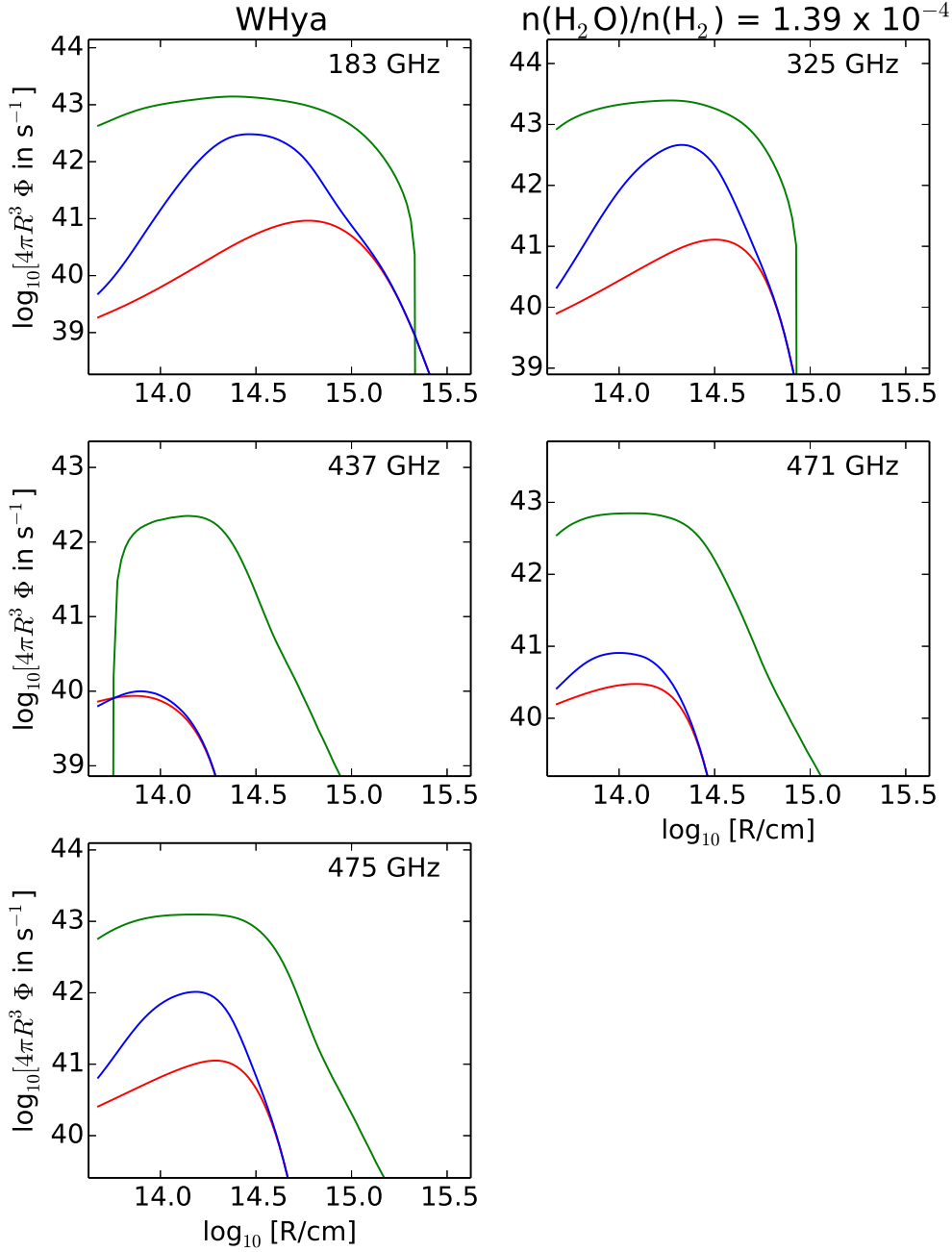}
\caption{Same as Figure 4, but for transitions of para-H$_2$O.}
\end{figure}

\begin{figure}
\includegraphics[width=15 cm]{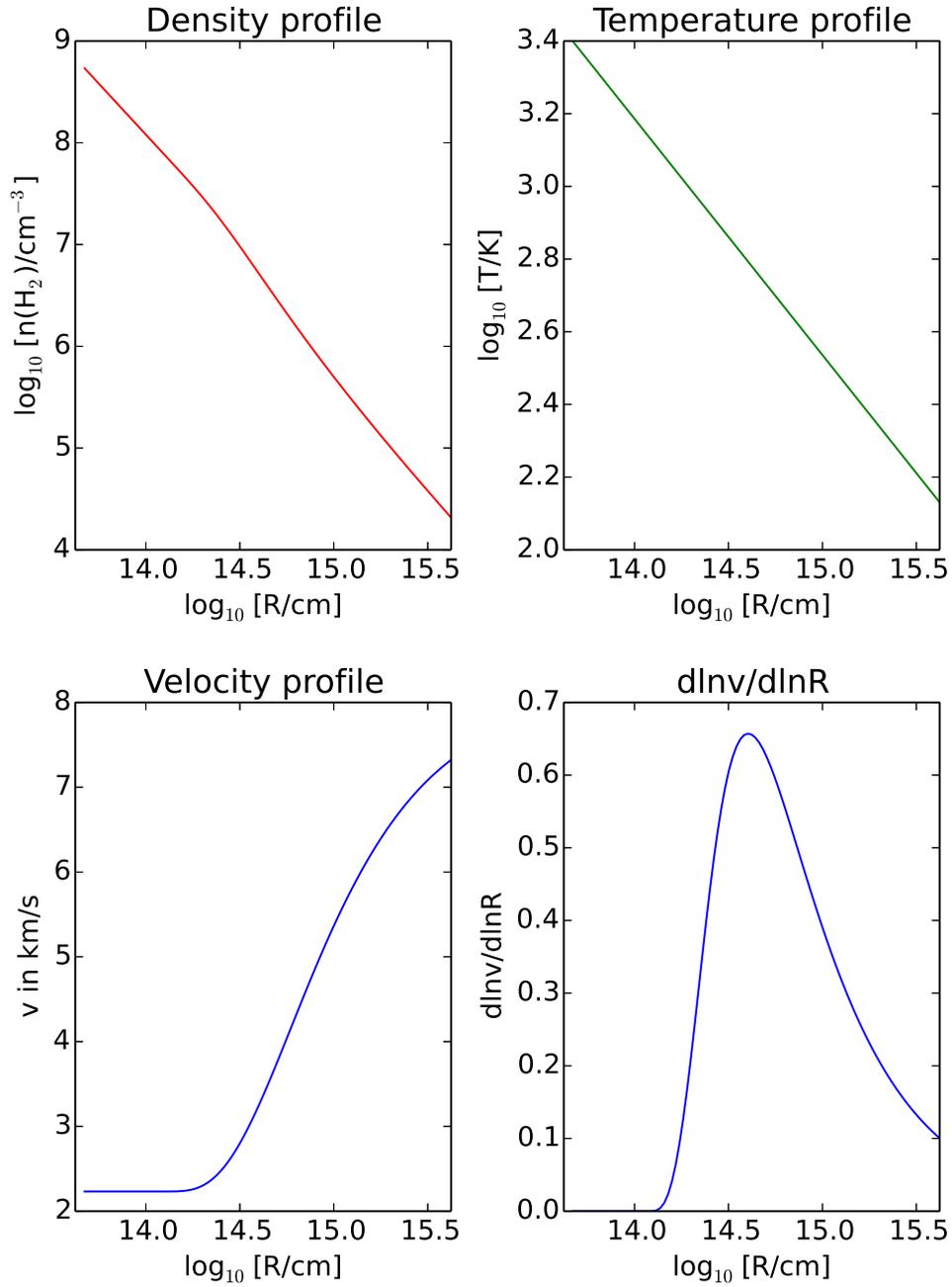}
\caption{Density, temperature, and velocity profiles assumed in the best-fit model for VY~CMa}
\end{figure}





In Figures 4 and 5, we show how the maser emissivity predicted in
the best-fit model obtained for W~Hya depends on the radial coordinate, $R$.
As in Paper 1, we present (red curve) the rate of spontaneous emission per logarithmic distance element, $4\pi R^3 \Phi_p({\rm spont})$, in units of photons s$^{-1}$; the total rate of photon emission per logarithmic distance element for saturated maser emission, $4\pi R^3 \Phi_p({\rm sat})$ (green); and our estimate of actual photon emission rate per logarithmic distance element, $4\pi R^3 \Phi_p$ (blue).  Thus, integrating $4\pi R^3 \Phi_p$ $d{\rm ln} R$ yields the predicted photon luminosity that is to be compared with the observed value.  Figure 4 shows the emissivity profiles for transitions of ortho-H$_2$O, and Figure 5 shows those for transitions of para-H$_2$O.  Figure 6 shows the density, temperature, and velocity profiles assumed in the model.  

Several noteworthy features are apparent in Figures 4 -- 6:  

\noindent (1) In the innermost part of the outflow ($R \le 1 - 3 \times 10^{14}$~cm) where the density is highest, $4\pi R^3 \Phi_p({\rm sat})$ tends to decrease (green curves).. This behavior demonstrates the phenomenon of maser quenching: as the density increases, collisions start to dominate radiative processes and ultimately drive the level populations to local thermodynamic equilibrium (LTE), eliminating any population inversion.  

\noindent (2) In the quenched regime, the optical depth is positive, and the estimated photon emission rate $\Phi_p$ (blue) drops below $\Phi_p({\rm spont})$ (red).  

\noindent (3) In the masing regime ($\Phi_p({\rm sat}) > 0$), by contrast, the estimated photon emission rate $\Phi_p$ may greatly exceed $\Phi_p({\rm spont})$.  However, even when $\Phi_p({\rm sat}) > 0$, a large ratio of $\Phi_p/\Phi_p({\rm spont})$ is not necessarily achieved: in the outer envelope beyond $R = {\rm few} \times 10^{14}\,\rm cm$, the gain is very small with $\Phi_p \sim \Phi_p({\rm spont})$.  In all cases, the photon emission rate $\Phi_p$ is always smaller than $\Phi_p({\rm spont})+\Phi_p({\rm sat}).$   

\noindent (4) The various transitions we have modeled have different predicted degrees of saturation.  In general, the 1.296~THz maser tends to be the most saturated, while the 22 GHz transition tends to have the smallest predicted $\Phi_p/\Phi_p({\rm sat})$ ratio.

\noindent (5) All the curves tend to drop at large $R$, primarily because the temperature and density are falling in the outer envelope and eventually become too low to excite the upper state of the transition.  As expected, this behavior is most marked for the highest-excitation transitions (at 321 GHz and 1.296 THz). 

\subsection{Dependence on the mass-loss rate}

In the second step of our analysis, we adopted the best-fit water abundances obtained in Section 4.2, and investigated the dependence of the line luminosities on the mass-loss rate.  Example results are shown for
VY CMa in Figures 7 (ortho-H$_2$O lines) and 8 (para-H$_2$O lines).  Colored curves show the predicted line luminosities (blue); and the luminosities that would be attained without maser amplification (red) and with saturated maser action (green).  Filled symbols indicate the estimated mass-loss rates (Table 1) and observed luminosities from Paper I (blue dots), M08 (red stars), Harwit et al.\ (2010; green dot), and the present work (black stars).  With the exception of the 437~GHz transition, all the observed luminosities are bracketed by the luminosities predicted without maser amplification and those predicted in the limit of saturated maser action.  As we found in Paper I, the 1.296 GHz transition can be regarded as a true water maser, with the model indicating -- for all sources where the transition was detected -- that the observed luminosity significantly exceeds that accounted for by spontaneous radiative decay. 

\begin{figure}
\includegraphics[width=14 cm]{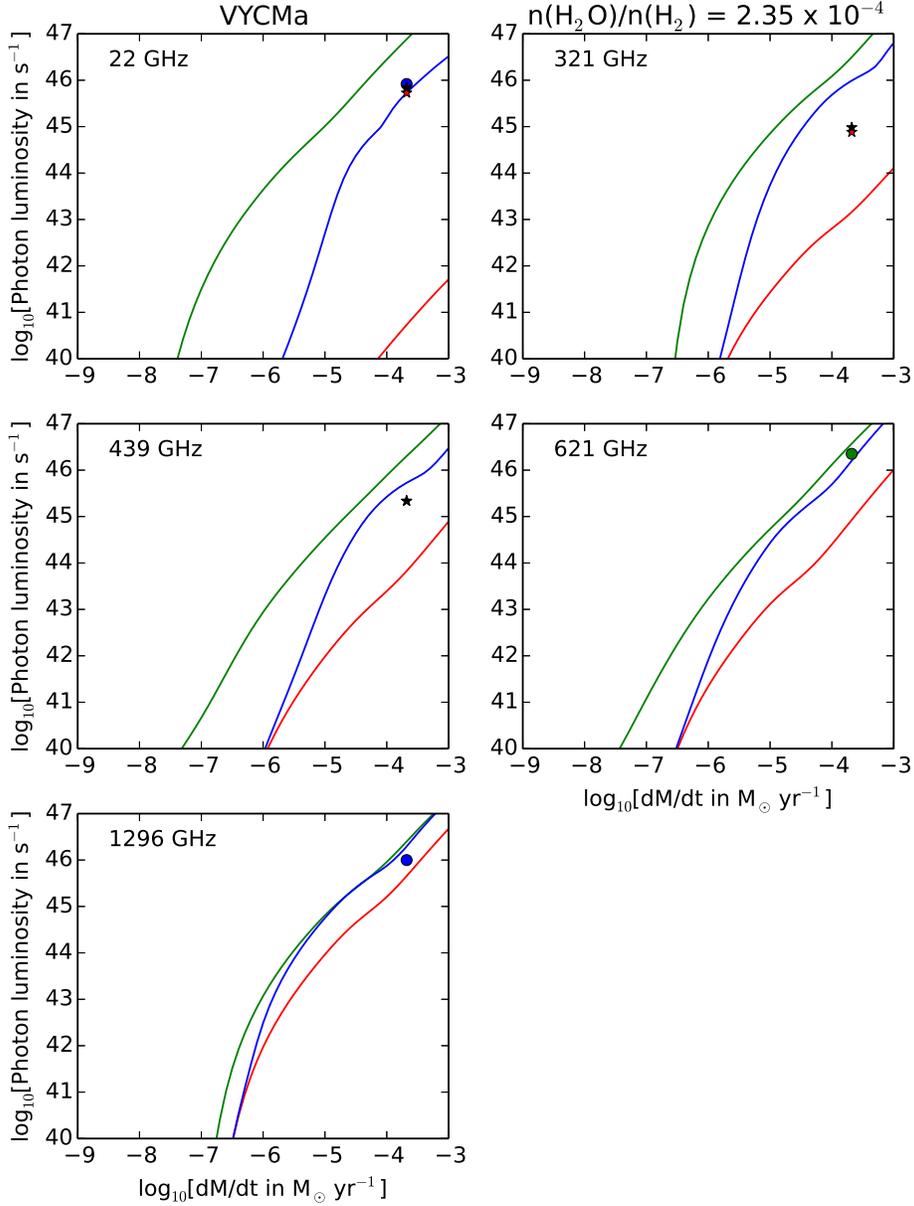}
\caption{Predicted photon luminosities from VY~CMa for transitions of ortho-H$_2$O, as a function of mass-loss rate, for the best-fit water abundances obtained in Section 4.2.  Colored curves show the predicted line luminosities (blue); and the luminosities that would be attained without maser amplification (red) and with saturated maser action (green). Filled symbols show the observed luminosities (see text for details).}
\end{figure}

\begin{figure}
\includegraphics[width=14 cm]{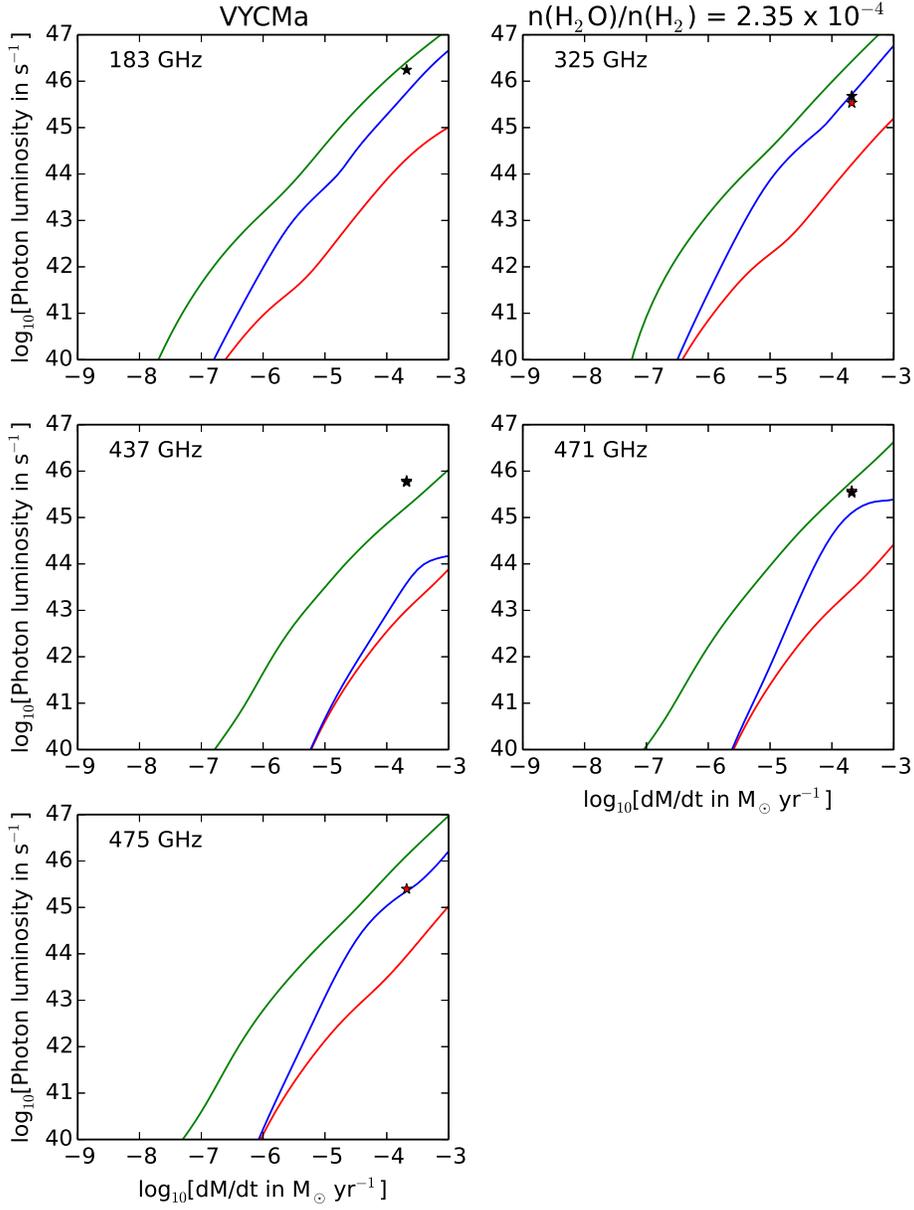}
\caption{Same as Figure 7, for transitions of para-H$_2$O}
\end{figure}

\begin{figure}
\includegraphics[width=8 cm]{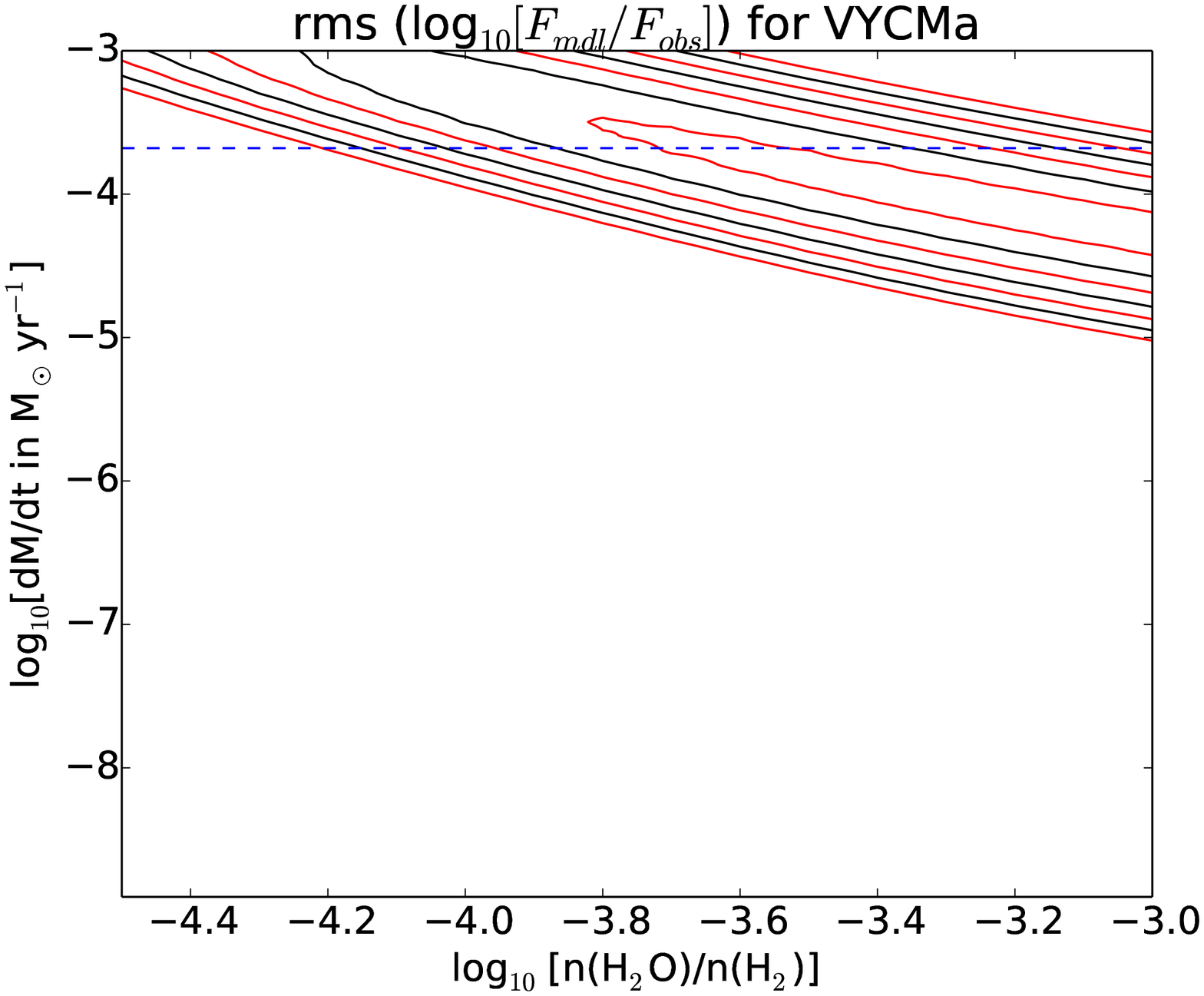}
\includegraphics[width=8 cm]{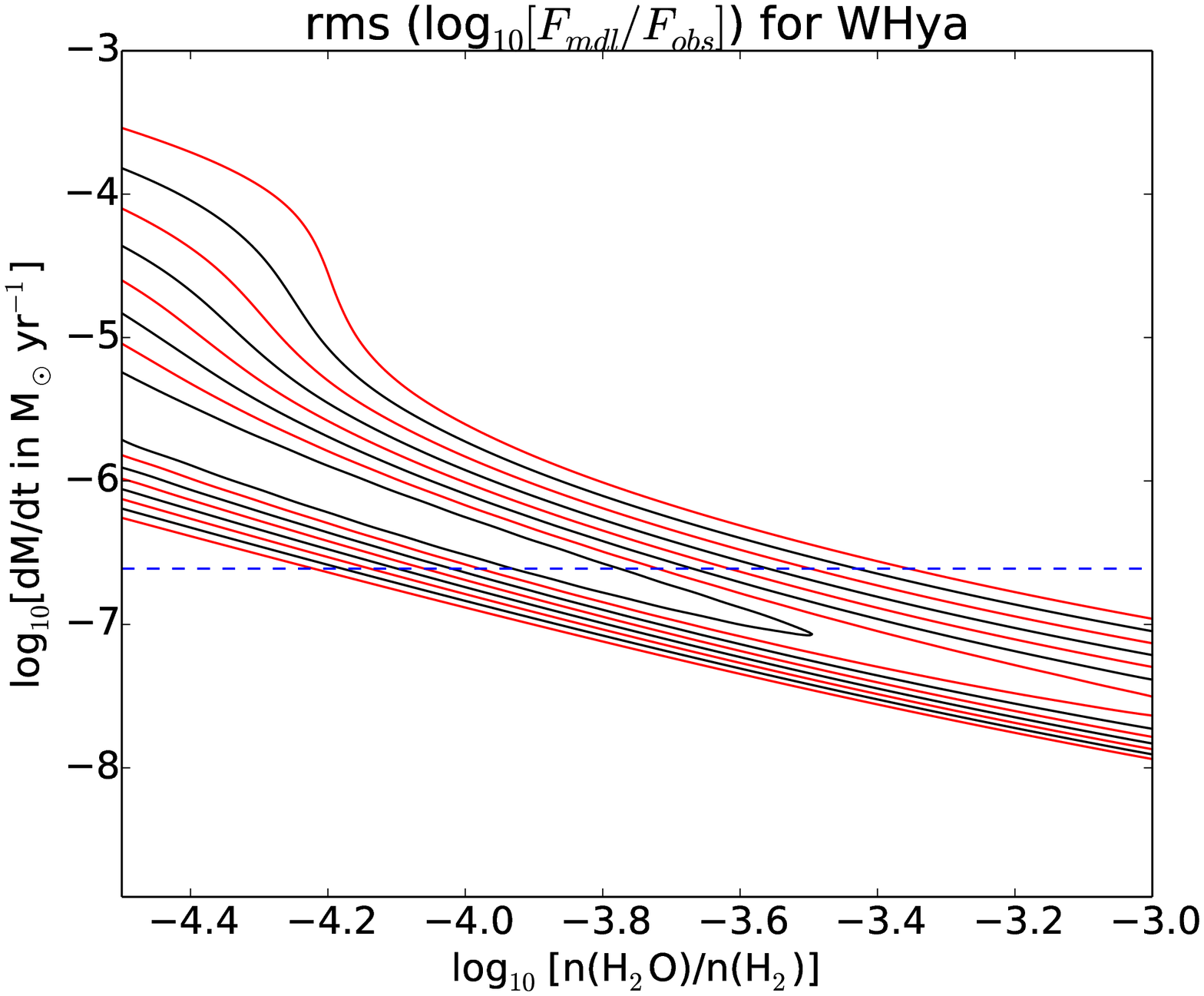}
\includegraphics[width=8 cm]{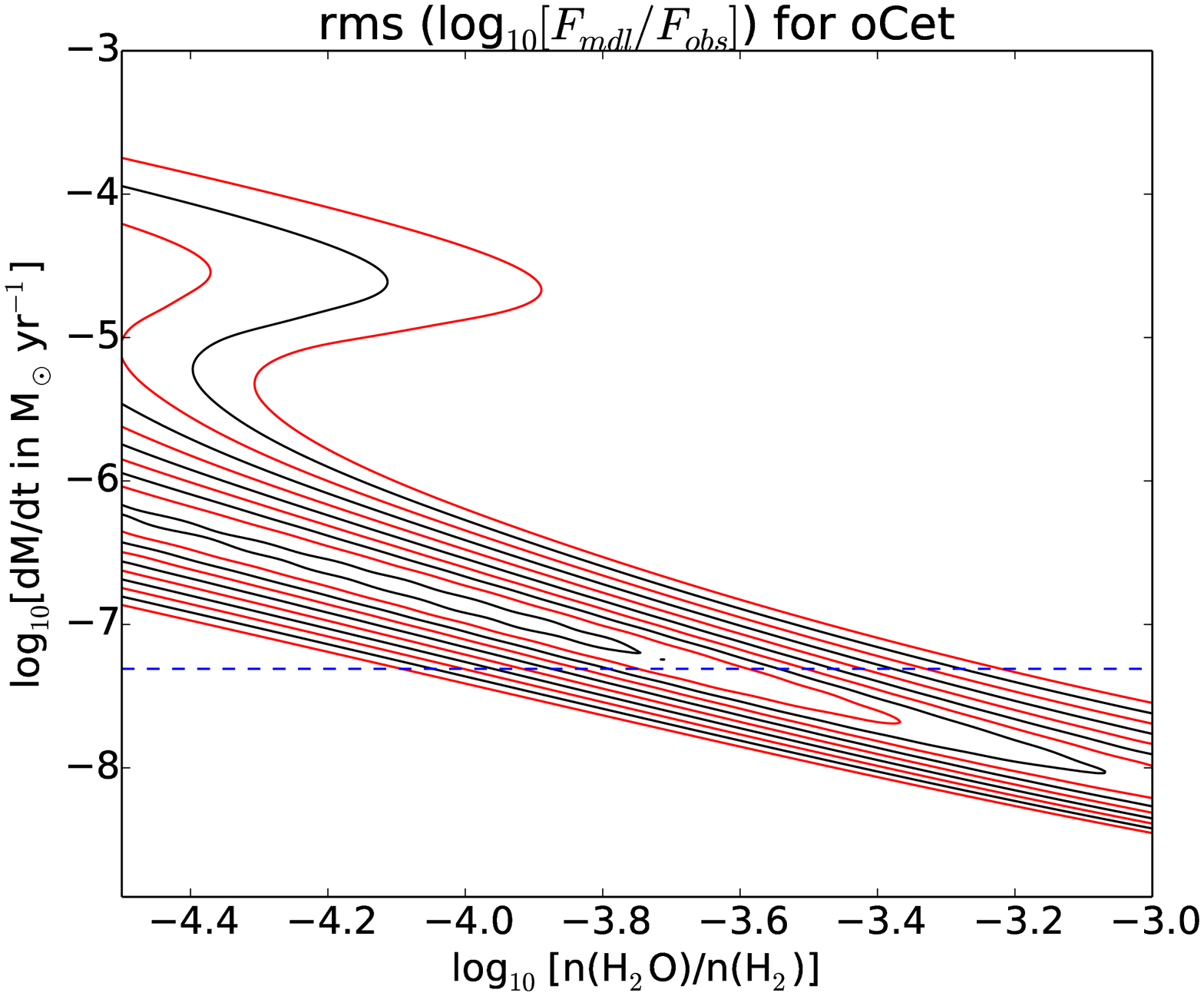}
\includegraphics[width=8 cm]{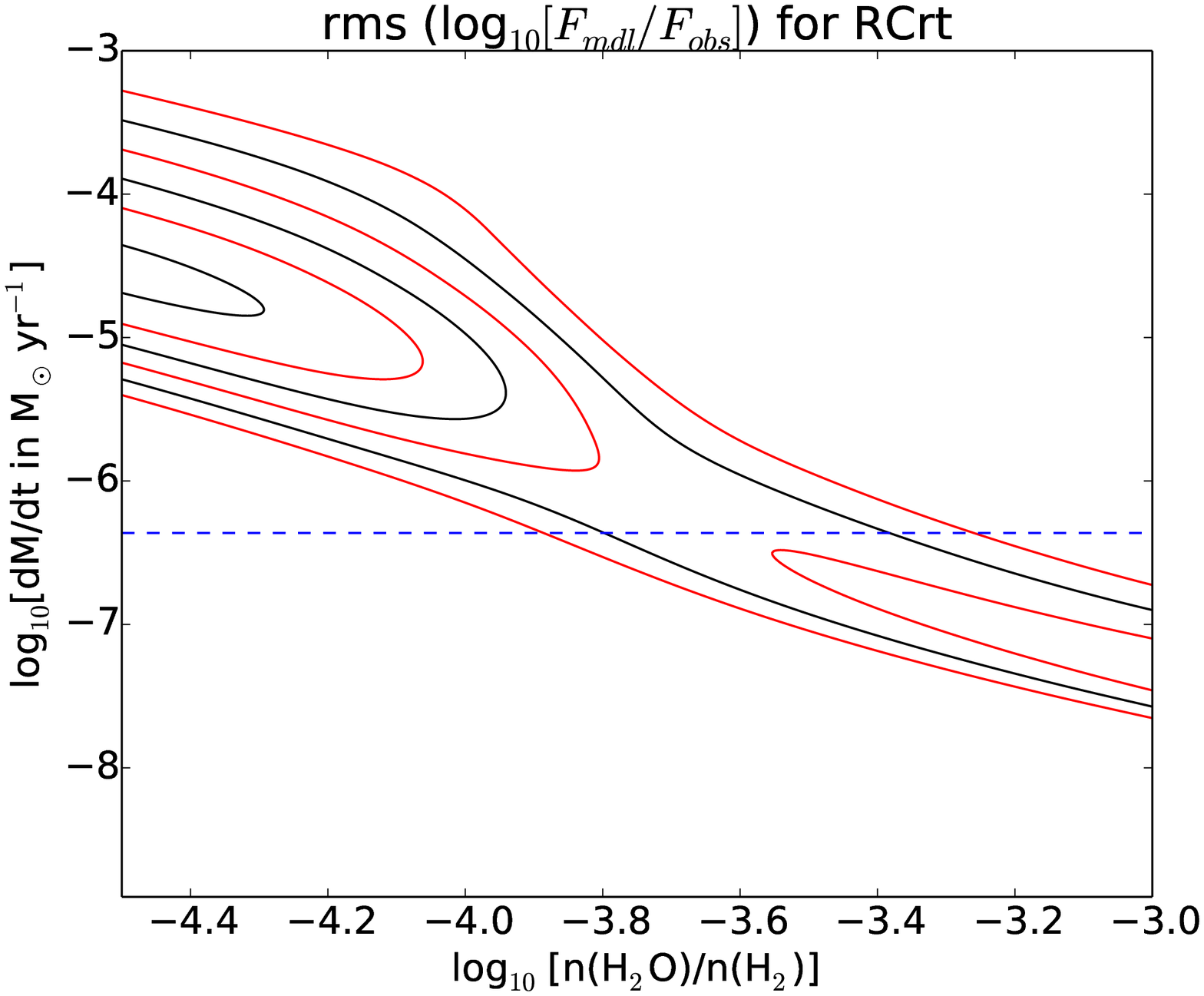}
\includegraphics[width=8 cm]{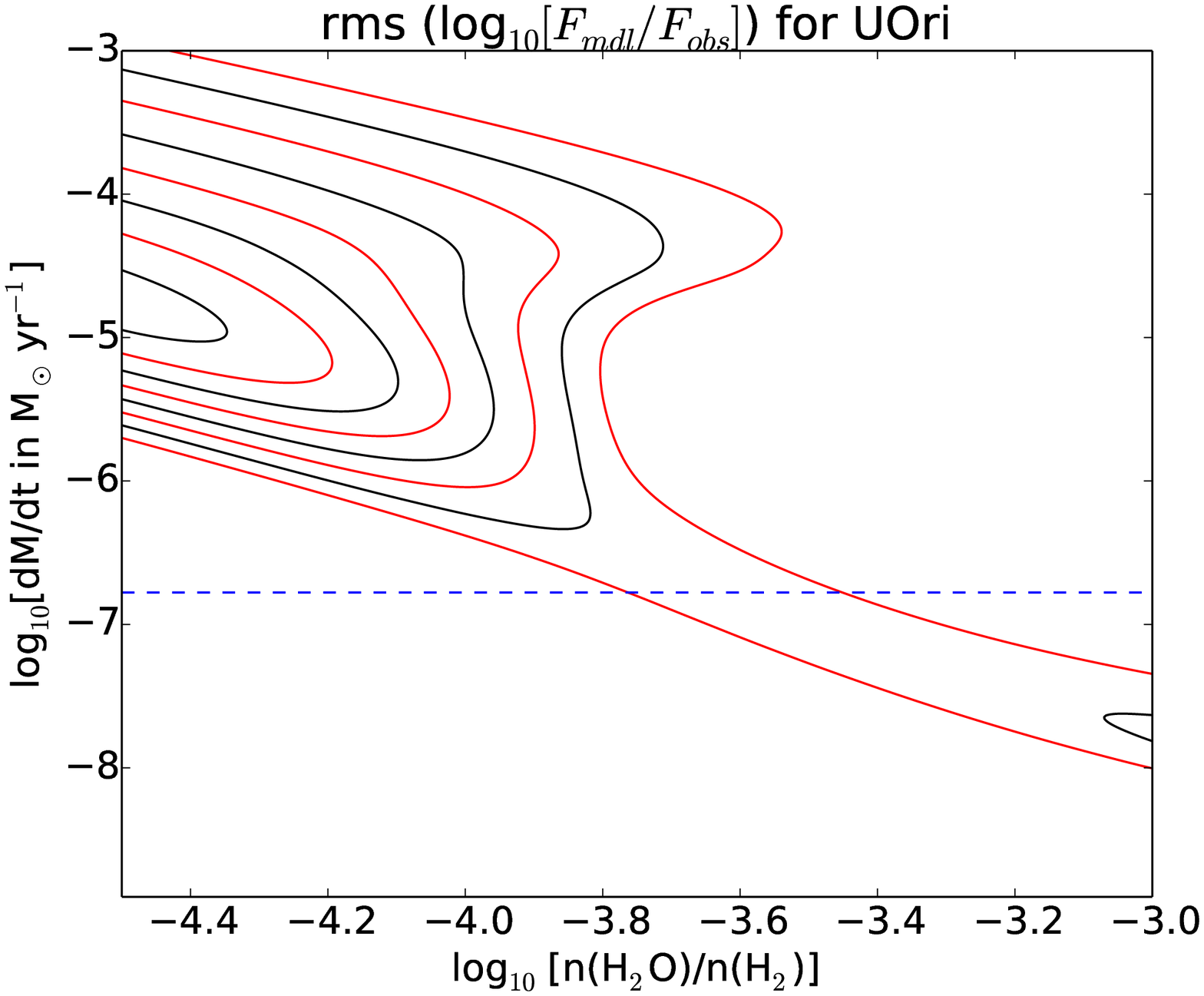}
\caption{Contours of ${\rm rms}\, ({\rm log}_{10}[F_{\rm mdl}/F_{\rm obs}])$ are plotted in the plane of $\dot M$ and $x_{\rm H2O}$.  The dotted blue lines indicate literature estimates of the mass-loss rates, based on CO line fluxes. The outermost contours show the locus where ${\rm rms} ({\rm log}_{10}[F_{\rm mdl}/F_{\rm obs}])=1.1$, with contours decreasing inwards in steps of 0.1}
\end{figure}

In the final step of our analysis, we relaxed all priors on the mass-loss rate and water abundance, and investigated the dependence of ${\rm rms} ({\rm log}_{10}[F_{\rm mdl}/F_{\rm obs}])$ on both $\dot M$ and $x_{\rm H2O}$.  The results are shown for each star in Figure 9, where contours of ${\rm rms} ({\rm log}_{10}[F_{\rm mdl}/F_{\rm obs}])$ are plotted in the plane of $\dot M$ and $x_{\rm H2O}$.  The outermost contours show the locus where ${\rm rms} ({\rm log}_{10}[F_{\rm mdl}/F_{\rm obs}])=1.1$, with contours decreasing inwards in steps of 0.1.  Not surprisingly, there is a considerable degeneracy between $\dot M$ and $x_{\rm H2O}$, with increasing values of $\dot M$ being largely compensated for by decreasing values of $x_{\rm H2O}$.

\subsection{Performance of this model with pure collisional pumping}

For three of the five sources listed in Table 4,
the best-fit models presented here succeed in fitting the observed 
fluxes for most transitions with a typical error of 0.21 - 0.48 dex, corresponding to 
a factor 1.6 -- 3, while for R~Crt and U Ori, the typical errors are 0.90 dex and 1.01 dex respectively, corresponding to much larger factors of 8 and 10.  These models adopt, as a prior, literature estimates of the mass-loss rates
that were derived by an entirely independent 
method based on CO line fluxes.   (Better fits for R~Crt and U Ori are possible (Figure 9), but only for assumed mass-loss rates that are much larger than those estimated from the CO line fluxes.)  The best-fit water abundances lie
in a remarkably narrow range, $1.4 \times 10^{-4}$ to $2.5  \times 10^{-4}$,
that is broadly consistent with the expectations for an oxygen-rich outflow and with estimates derived from {\it Herschel} observations of non-masing water transitions (Maercker et al.\ 2016).   {As noted in Section 4.1 above, the neglect of radiative excitation in our model may lead to overestimates of the maser line luminosities that are predicted for a given mass-loss rate, particularly for the least saturated transitions (e.g. the 22 GHz line).  Thus the water abundances we derived -- if anything -- may be slightly underestimated.  The importance of this shortcoming of the model is mitigated by the fact that the least saturated transitions, for which the line luminosities are affected most by radiation excitation by dust continuum radiation, are precisely those transitions for which the line luminosity is also most strongly dependent on the water abundance (Figure 3).}

\begin{figure}
\includegraphics[width=15 cm]{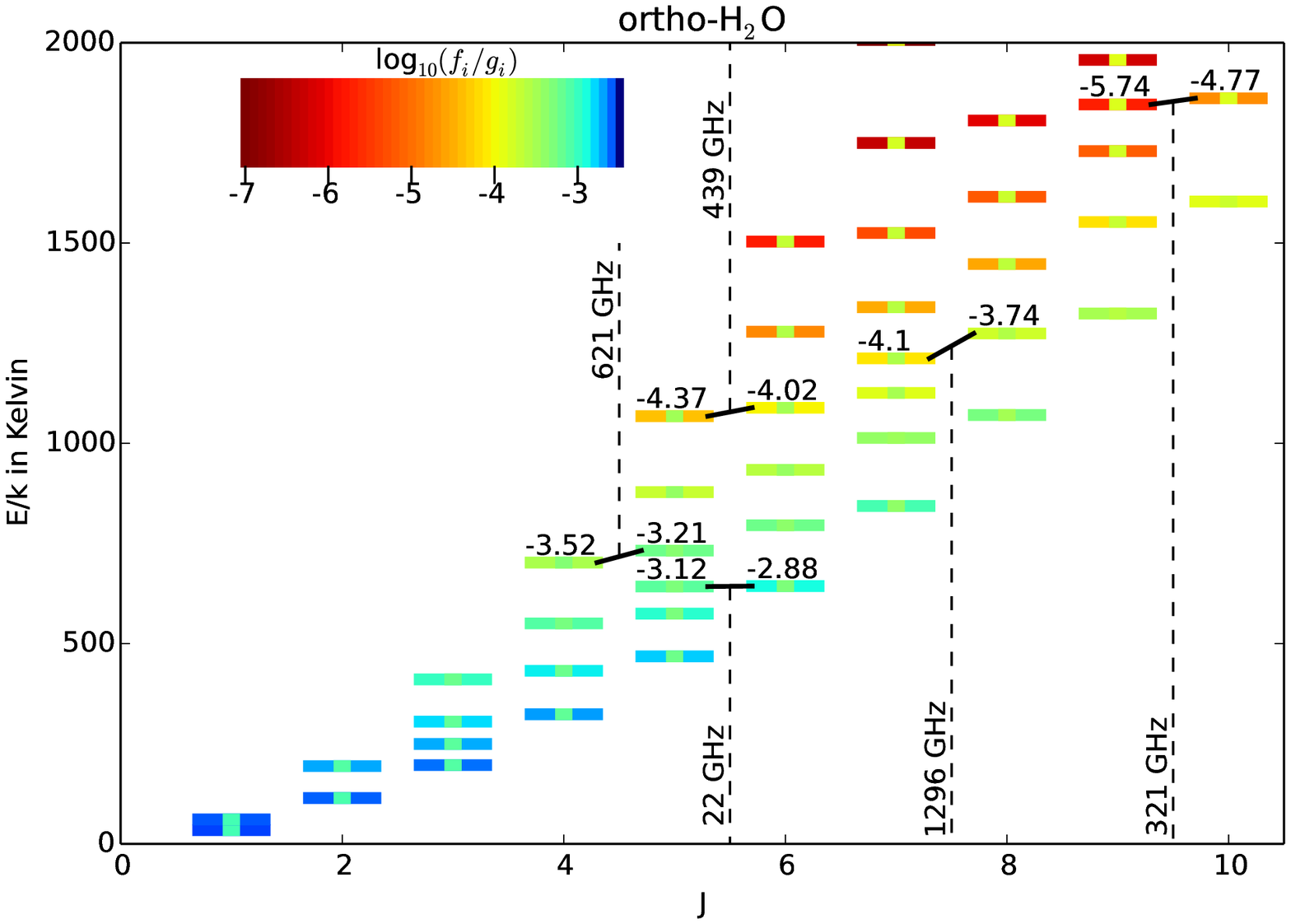}
\includegraphics[width=15 cm]{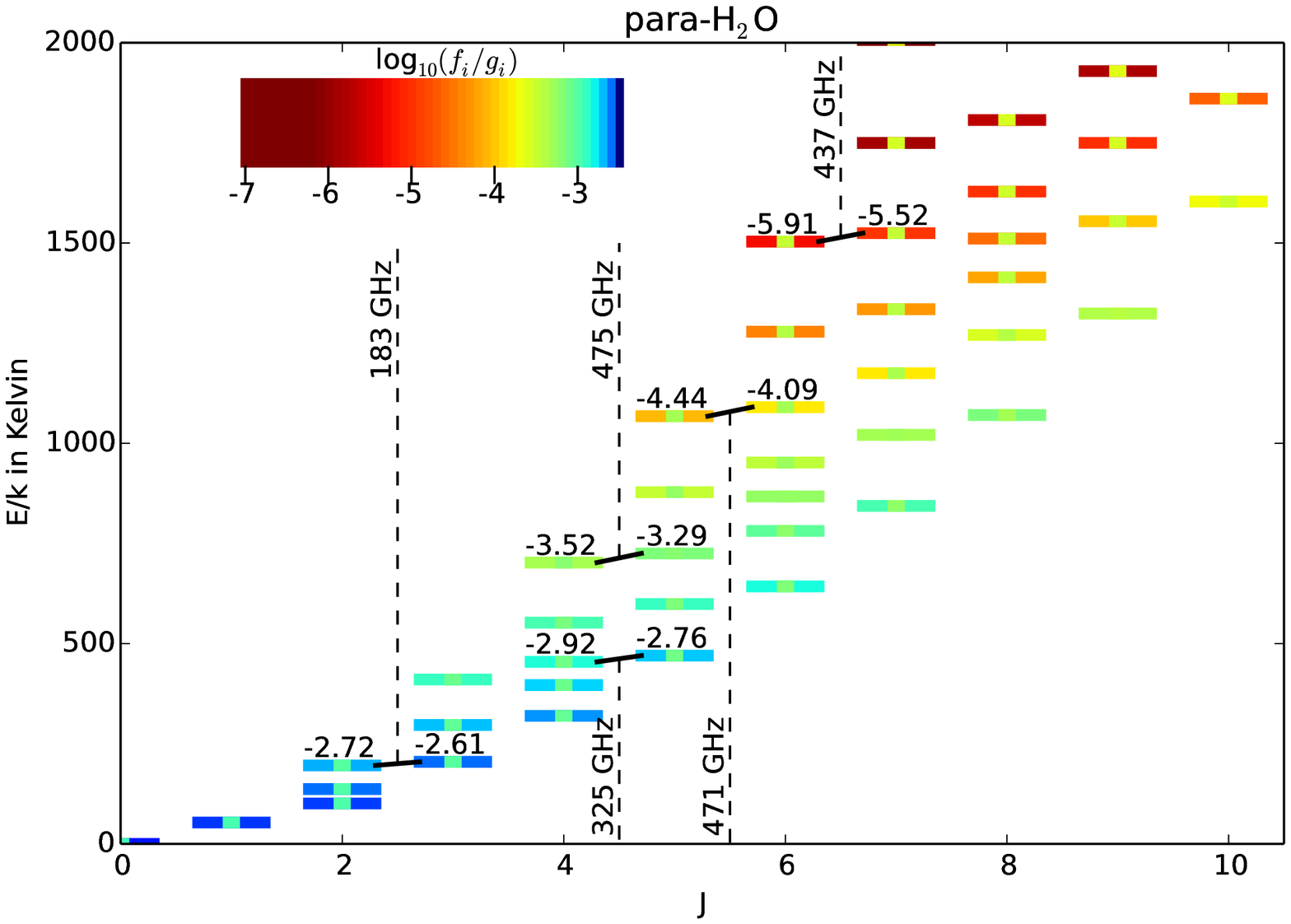}
\caption{Energy level diagrams for ortho-H$_2$O (top) and para-H$_2$O (bottom), with the color coding representing the LTE and non-LTE level populations for a typical circumstellar envelope (see the text for details).} 
\end{figure}

All ten maser transitions observed toward VY CMa are predicted to be inverted under the conditions prevalent in the outflow, although the luminosity of the 437~GHz transition is severely 
underpredicted.  As discussed in NM91, population inversions are typically 
predicted for submillimeter transitions with $\Delta J = -1$, provided that the density
is not so high as to force the level populations very close to LTE.   This behavior can 
be understood from Figure 10, which shows the arrangement of the
energy levels for both ortho- and para-H$_2$O.  For each value of $J$, there is a ``ladder" 
of states with different energies, each possessing a different projection of $J$ onto the 
principal axes of the molecule.  In Figure 10, each state $i$ is color-coded according to
its steady-state population, $f_i/g_i$, where $f_i$ is the fraction of water molecules in 
state $i$, $g_i=(2I+1)(2J+1)$ is the degeneracy, and $I$ is the total 
nuclear spin (0 for para-H$_2$O and 1 for ortho-H$_2$O).  The criterion for maser action
is that $f_i/g_i$ is larger for the upper state than for the lower.

The results shown here apply for conditions typical of circumstellar outflows: $T=1000$~K, 
$n({\rm H}_2) = 10^7\,\rm cm^{-3}$, and $n(\rm{H_2O})/(dv/dr) = 10^{17}\,\rm cm^{-2}$ 
per km/s.  Colorbars in each panel (upper left) indicate the color-coding, and 
black lines show the ten masing transitions listed in Table 1. 
The color of the outer parts of 
each bar shows the non-LTE level populations, while the color
at the center shows the LTE populations at 1000~K.  The numbers above the upper and lower
states of each listed maser transition indicate the non-LTE values of log$_{10}(f_i/g_i)$. 

In LTE, of course, $f_i/g_i$ depends only on the energy of the state,
and there are no population inversions.  But for the 
conditions considered here, there are significant departures from LTE.  There is a marked 
trend for the populations to become increasingly subthermal as the energy increases 
within a given ladder.  The populations are largest (and may indeed exceed the LTE values)
within the states at the bottom of each ladder, a set of states known as the ``backbone."
For states that are close to a {\it given energy}, the trend is therefore
for $f_i/g_i$ to increase with $J$, leading to population inversions in the $\Delta J = -1$
transitions.

While the water abundances derived from our model draw a consistent picture for all five sources, the model shows significant shortcomings, yielding large errors in the predicted line ratios for two of the five sources.  This behavior is perhaps unsurprising, given the simplicity of the model and the exponential nature of maser amplification.  Moreover, as has been noted recently by Bergman \& Humphreys (2020; hereafter BH20), smooth flow models -- such as the one presented here -- tend to predict radial beaming that would lead to double-peaked line profiles that are not typically observed.  Given the large velocity gradient (LVG) approximation that we adopted in the present study, maser emission is beamed radially whenever $d{\rm ln}v/d{\rm ln}R$ is less than unity, as is typically the case (see Figure 6).  BH20 suggested that strong shocks in the inner envelope might be implicated by this discrepancy, although definitive conclusions will have to await the development of 
circumstellar outflow models that include such shocks.

\section{Summary}

\noindent (1) Using the GREAT instrument on SOFIA, we have detected emission in the 1.296411 THz $8_{27}-7_{34}$ maser transition of water vapor toward two oxygen-rich evolved stars, omicron Ceti (Mira) and R~Crateris, and obtained an upper limit on the 1.296~THz line emission from U Orionis.  Combined with the earlier detections reported in Paper I, this brings the number of evolved stars with detected terahertz maser emission to five.

\noindent (2) Toward o~Cet, R~Crt and U Ori, and toward the red supergiant star VY Canis Majorae from which 1.296~THz line emission was reported in Paper I, we have also observed several lower-frequency water maser transitions.  The comprise the 22 GHz transition, observed using the Effelsberg 100-m telescope; and the 183, 321, 325, 437, 439, 471, and 475 GHz transitions, observed using the APEX 12~m telescope 

\noindent (3) We have used a simple model to analyse the multi-transition data thereby obtained. In this model, the maser transitions are excited by collisional excitation.  The arrangement of the energy states leads naturally to population inversions for all the $\Delta J = -1$ transitions that are observed to mase.  However, the 437~GHz maser luminosity is greatly underpredicted by the model, a long-standing puzzle.

\noindent (4) Adopting, as a prior, independent literature estimates of the mass-loss-rates in these four sources and in W~Hydrae, we infer water abundances in a remarkably narrow range: $n({\rm H_2O})/n({\rm H_2}) = 1.4 - 2.5 \times 10^{-4}$.  For $\it o$ Cet, VY CMa, and W Hya, the model is successful in predicting the maser line fluxes to within a typical factor $\sim 1.6 - 3$.  For R~Crt and U Ori, the model is less successful, with typical line flux predictions lying an order of magnitude above or below the observations; such discrepancies are perhaps unsurprising given the exponential nature of maser amplification.

\noindent (5) As we found for W~Hya in Paper I, the 1.296 GHz transition can be regarded as a true water maser, with the model indicating  -- for all sources where the transition was detected -- that the observed luminosity significantly exceeds that accounted for by spontaneous radiative decay.  It has the highest frequency of any water transition known to dominated by stimulated emission.
\\

\begin{acknowledgements}
Based on observations made with the NASA/DLR Stratospheric Observatory for Infrared Astronomy, the 100 m radio telescope
of the MPIfR in Effelsberg, and the 12 m APEX submillimeter telescope. SOFIA Science Mission Operations are conducted jointly by the Universities Space Research Association, Inc., under NASA contract NAS2-97001, and the Deutsches SOFIA Institut under DLR contract 50 OK 0901.   The development and operation of 4GREAT was financed by resources from the MPI f\"ur Radioastronomie, Bonn and the Universit\"at zu K\"oln; and by the Deutsche Forschungsgemeinschaft (DFG) within the grant for the Collaborative Research Center 956 as well as by the Federal Ministry of Economics and Energy (BMWI) via the German Space Agency (DLR) under Grants 50 OK 1102, 50 OK 1103 and 50 OK 110.  APEX is a collaboration between the Max-Planck-Instit\"ut f\"ur Radioastronomie, the European Southern Observatory, and the Onsala Space Observatory.
This research was supported by USRA through a grant for SOFIA Program 06-0015.  We gratefully acknowledge the outstanding support provided by the SOFIA Operations Team and the GREAT Instrument Team.  We thank Dirk Muders for clarifying aspects of APEX data calibration issues.

\end{acknowledgements}

{}


\begin{deluxetable}{lcccc}
\tablewidth{0pt}
\tabletypesize{\scriptsize}
\tablecaption{Sources observed in the present study} 
\tablehead{Source: \phantom{000} & \phantom{000} o~Cet \phantom{000}& \phantom{000}R~Crt \phantom{000} & \phantom{000} U~Ori \phantom{000} & \phantom{000} VY~CMa \phantom{000} }
\startdata
  \multicolumn{3}{l}{\underbar{Source parameters}} \\
  \phantom{000000}RA (J2000) & 02h 19m $20 \fs 79$ & 11h 00m $33 \fs 85$ & 05h 55m $49 \fs 17$ & 07h 22m 58\fs 33\\  
  \phantom{000000}Dec.\ (J2000)) & --02$^o$ $58^{\prime}$ $39 \farcs 5$ 
   & --18$^o$ $19^{\prime}$ $29 \farcs 6$ 
   & +20$^o$ $10^{\prime}$ $30 \farcs 69$ 
   & --25$^o$ $46^{\prime}$ $03 \farcs 24$ \\
  \phantom{000000}Distance$^a$ (pc) & $93\pm 10^b$  & $238 \pm 12$ & $228 \pm 18$ & $1200 \pm 115^c$ \\
  \phantom{000000}Spectral type$^d$ 				& M5e-M9e & M7 & M6e-M9.5e & M5eIbp(C6,3)  \\
  \phantom{000000}Variability type$^d$ 		    & Mira & SRb & Mira & RSG \\
  \phantom{000000}Variability period$^d$ (days) 	&  332 & 160 & 368 & 956\\  
  \phantom{000000}Systemic LSR velocity$^e$ (km s$^{-1}$) 	& $46.6 \pm 0.2$& $11.2 \pm 0.7$  & $-38.1 \pm 1.3$ & $17.6 \pm 1.5$\\
  \phantom{000000}Expansion velocity$^e$ (km s$^{-1}$) 		& \phantom{0}$2.4 \pm 0.4^f$	& $10.8 \pm 0.7$  & \phantom{--0}$7.5 \pm 1.2$ & $44$\\
  \phantom{000000}Mass-loss rate$^g$ ($10^{-7}\,{\rm M}_\odot \rm \, yr^{-1}$) & 0.49$^f$ & 4.34 & 1.67 & 2090 \\
\multicolumn{3}{l}{\underbar{Observational parameters}} \\  
  \phantom{000000}Date of SOFIA observation & 2018 Dec 12 & 2018 Dec 12 & 2018 Dec 11, 12 & (h) \\
  \phantom{000000}Stellar phase$^i$ & 0.02 & ... & 0.64 & ...\\ 
  \phantom{000000}Date of Effelsberg observation & 2018 Dec 11 & 2018 Dec 11 & 2018 Dec 11 & 2018 Dec 11 \\
  \phantom{000000}Date of APEX observations & 2018 11/30 - 12/6 & 2018 11/30 - 12/6 & 2018 11/30 - 12/6 & 2018 11/30 - 12/6\\
  \phantom{000000}Source velocity w.r.t.\ SOFIA (km s$^{-1}$) &  $78$ & $-9$ &  $-30$ & (h)\\
  \phantom{000000}SOFIA integration time (min, on-source) & 12.4 & 8.0 & 10.2 & (h) \\  
\hline
\multicolumn{5}{l}{$^{a}$Weighted average of {\it Hipparcos} and {\it Gaia} DR2 results, where available, except for VY~CMa}\\
\multicolumn{5}{l}{$^{b}$Where given, errors are 1~$\sigma$ statistical errors }\\
\multicolumn{5}{l}{$^{c}$Zhang et al.\ (2012)}\\
\multicolumn{5}{l}{$^{d}$General Catalogue of Variable Stars, v5.1 (Samus et al.\ 2017)} \\
\multicolumn{5}{l}{$^{e}$As determined from observations of thermal CO emission (Knapp et al.\ 1998; Matsuura et al.\ 2014 for VY~CMa)} \\
\multicolumn{5}{l}{$^{f}$Values applying to the slow wind.  {The fast wind has an expansion velocity of 6.7~km$ \, \rm s^{-1}$ (K98), but does not}}  \\
\multicolumn{5}{l}{{emit detectable water maser emission (see the text).}} \\
\multicolumn{5}{l}{$^{g}$Estimated from thermal CO emission (Knapp et al.\ 1998; Matsuura et al.\ 2014 for VY~CMa) and scaled by $D^2$ (see text)} \\
\multicolumn{5}{l}{$^{h}$Not observed in Cycle 6; see Paper I for details of Cycle 4 observations} \\
\multicolumn{5}{l}{$^{i}$Waagen et al. (Journal of AAVSO, in preparation, www.aavso.org/maxmin)} \\
\enddata
\end{deluxetable}

\begin{deluxetable}{ccccc}
\label{lines}
\tablewidth{0pt}
\tablecaption{List of water maser transitions} 
\tablehead{Frequency  	& Transition 	& Spin symmetry 	& Upper state & Jy/K\\
		  (GHz) 		&   			& 					& energy / k}
\startdata
22.23508		& $6_{16}-5_{23}$	& ortho 		&	644~K		\\
183.3101	& $3_{13}-2_{20}$	& para			&	205~K	& 37	\\
321.2257	& $10_{29}-9_{36}$	& ortho			&	1861~K	& 41	\\
325.1529	& $5_{15}-4_{22}$	& para			&	470~K	& 41	\\
437.3467	& $7_{53}-6_{60}$	& para			&	1525~K	& 46	\\
439.1508	& $6_{43}-5_{50}$	& ortho			&	1089~K	& 46	\\
470.8889	& $6_{42}-5_{51}$	& para			&	1090~K	& 48	\\
474.6891	& $5_{33}-4_{40}$   & para			&	725~K	& 48	\\
620.7010	& $5_{32}-4_{41}$ 	& ortho			&	732~K		\\	
1296.4110	& $8_{27}-7_{34}$   & ortho			&	1274~K		\\
\enddata
\tablecomments{Columns are, left to right, for all observed lines, the rest frequencies, quantum numbers, spin symmetry, upper state energy above the ground state. The final column lists, for the lines observed with the APEX telescope, the antenna temperature to Jy conversion factor (see text). }
\end{deluxetable}

\begin{deluxetable}{cccccc}
\tablewidth{0pt}
\tabletypesize{\scriptsize}
\tablecaption{Results of observations} 
\tablehead{Line  & $\int S d\nu$ 	& $v$-range 	& $S_p$ & $v_p$ 		& $L_\nu$ \\
		(GHz)	     & (Jy km~s$^{-1}$) & (km~s$^{-1}$) & (Jy) 	& (km~s$^{-1}$) & (s$^{-1}$)}
\startdata
\multicolumn{6}{c}{\bf --- o Cet ---} \\
22		& 5.1 $\,\pm\,$0.02	& [42,50] 	&	3.8	&	46.89 & $2.7 \times 10^{40}$ 	\\
183     & 271 $\,\pm\,$5        & [45,48]   &  143  &  46.9     & $1.4 \times 10^{42}$    \\ 
321     &  10 $\,\pm\,$3        & \phantom{*}[45,48]*  & $\approx 6$ &     --  & $5.2 \times 10^{40}$\\ 
325     & 127 $\,\pm\,$12       & [45,48]   &  116  &  47.0      &  $6.6 \times 10^{41}$ \\ 
437     & 83\ $\,\pm\, $8        & [44,53]   &  $\approx13$  &  $\approx49$  & $4.3 \times 10^{41}$\\
439     & $< 54$                & \phantom{*}[45,48]*  &     $<44$  &   --  &  --  \\
471     &$<49$                  & \phantom{*}[45,48]*  &$<30$       &--     &  --  \\
475     &\multicolumn{5}{c}{--- not observed ---}\\
1296    & 1730 $\,\pm\,$160    &  [43,49]   & $400$      & $46$ & $9.0 \times 10^{42}$ \\%
\hline
\multicolumn{6}{c}{\bf --- R Crt ---} \\
22      & 1675$\,\pm\,$1     & [3,17]       &   521 & 9.1 & $5.7 \times 10^{43}$ \\
183     &2821$\,\pm\,$7     & [1,23]        &   303 & & $9.6 \times 10^{43}$ \\
321     &39$\,\pm\,$2       & [3,$\approx25$] &  5  &  10.6 & $1.3 \times 10^{42}$\\
325     &794$\,\pm\,$7      & [4,17]        &161  & 9 & $2.7 \times 10^{43}$\\
437     &$<21$              & \phantom{*}[0,23]*       &$<7$&    -- & -- \\
439     &24$\,\pm\,$12      & \phantom{*}[0,23]*      &$<12$&   --  & $8.2 \times 10^{41}$\\
471     &25$\,\pm\,$10      & \phantom{*}[0,23]*      &$<7$&   --  & $8.5 \times 10^{41}$\\
475     &$<914$             & \phantom{*}[0,23]*       &$<22$& -- & -- \\
1296	&$\approx 1200 \,\pm\,499$	& \phantom{*}[0,23]*		&   $\approx 230$	&    $\approx 8 $	& $1.1 \times 10^{44}$	\\
\hline
\multicolumn{6}{c}{\bf --- U Ori ---} \\
22      & 10.8$\,\pm\,$0.1    & [--42,--36]   &    6.5&   --38.4 & $3.4 \times 10^{41}$\\
183     &645$\,\pm\,$19     & [--45,--45]  &$\approx 115$        & & $2.0 \times 10^{43}$ \\
321     &$6\,\pm\,$2        & [--41,--39]  &$\approx5$ &--40 & $1.9 \times 10^{41}$\\
325     &182$\,\pm\,$33    &  [--44,--36]  &$\approx 64$  &-- & $5.7 \times 10^{42}$\\
437     &$<62$             & \phantom{*}[--46,--35]*   & $<29$ & -- & --\\
439     &$< 61$             & \phantom{*}[--46,--35]* & $<29$  &-- & --\\
471     & \multicolumn{5}{c}{--- not observed ---}\\
475     &$<210$           & \phantom{*}[--46,--35]*  & $< 73$ & -- & --\\ 
1296    &$<695$         & \phantom{*}[--46,--35]*  & $<280$ & --   & --\\
\\

\multicolumn{6}{c}{\bf --- VY CMa ---} \\
22	    & 7675$\,\pm\,$1   & [--6,48] &1244 & 12.8 & $6.8 \times 10^{45}$\\
183	    & 19622$\,\pm\,$16 &[--10,63]  & 692 & 0.1 & $1.7 \times 10^{46}$\\
321	    & 1080$\,\pm\,$6   &[6 36]     & 312 & 18 & $9.6 \times 10^{44}$\\
325  	& 5362$\,\pm\,$61  & [--9,51]  & 284 & 13 & $4.8 \times 10^{45}$\\
437	    &  6493$\,\pm\,$22 & [ --1,44] & 509 & 29 & $5.8 \times 10^{45}$\\
439	    &  2444$\,\pm\,$47 &  [--1,36] & 161 & 14 & $2.1 \times 10^{45}$\\
471	    &  3835$\,\pm\,$28 & [--7,41]  & 190 & 22 & $3.4 \times 10^{45}$\\
475	    & \multicolumn{5}{c}{--- not observed ---}\\
1296    & \multicolumn{5}{c}{--- not observed ---}\\
\enddata
\tablecomments{Columns are, left to right, line frequency (rounded GHz), velocity integrated flux density, LSR velocity range showing emission, flux density and LSR velocity of strongest emission and isotropic photon luminosity. The errors in  $\int S d\nu$ are purely statistical values and do not include the generally larger calibration uncertainties (see text). Upper limits for  $\int S d\nu$ 	and $L_\nu$ are three times the value calculated from the rms noise level considering a line's indicated $v$-range. The latter was determined by visual inspection using CLASS analysis tools. It has an uncertainly of $\approx 1$~km~s$^{-1}$, which may be larger for lines with poor signal to noise ratio. An asterisk marks entries for which the $v$-range was fixed to the listed value, which is representative of those of other lines.
}

\end{deluxetable}

\begin{deluxetable}{lccccc}
\tablewidth{0pt}
\tabletypesize{\scriptsize}
\tablecaption{Model parameters and results} 
\tablehead{Source: \phantom{000} & \phantom{000} o~Cet \phantom{000} & \phantom{000}R~Crt \phantom{000} & \phantom{000} U~Ori \phantom{000} & \phantom{000} VY~CMa \phantom{000}  & \phantom{000} W~Hya \phantom{000}}
\startdata
  \multicolumn{6}{l}{\underbar{Parameters for the assumed temperature profile}, $T = T_* (R/R_*)^{-\epsilon}$}  \\
$R_*$/(10$^{13}$ cm)    & 2.31$^a$ & 3.91$^b$ & 3.15$^a$ & 14.4$^c$ & 4.72$^d$ \\
$T_*$/K			        & 2800$^a$ & 2800$^b$ & 2320$^a$ & 2800$^c$ & 2500$^d$ \\
$\epsilon$		        & 0.65$^e$ & 0.65$^e$ & 0.65$^e$ &	(f)		& 0.65$^d$ \\
$R_{\rm in}/R_*$        & 2.5$^e$  & 2.5$^e$  &  2.5$^e$ &	8.9$^c$ & 2.5$^e$  \\
  \multicolumn{6}{l}{\underbar{Parameters for the assumed velocity profile}, $v = v_0 + (v_\infty - v_0)(1 - R_{\rm in}/R)^\beta$} \\
$v_0$ / km~s$^{-1}$		&  (g)     &  (g)     &	 (g)	 &	4.0$^c$ & (g)	   \\
$\beta$					& 1.5$^h$ & 1.5$^h$   & 1.5$^h$ &	0.2$^c$ & 5.0$^d$. \\
$v_\infty$ / km~s$^{-1}$ (i) &	2.4	& 10.8	&	7.5		&	44	&	8.1		\\
\underbar{Density law$^j$}  & $n \propto 1/(r^2v)$  &$n \propto 1/(r^2v)$ 	&$n \propto 1/(r^2v)$ 	& $n \propto 1/r^{2}$ 	&	$n \propto 1/(r^2v)$ \\
  \multicolumn{6}{l}{\underbar{Characteristics of the best-fit model obtained with the $\dot M$ in Table 1}} \\
${\rm rms} ({\rm log}_{10}[F_{\rm mdl}/F_{\rm obs}])$ & 0.21	& 0.91	&	1.04		&	0.48		&	0.33		\\
$n({\rm H_2O})/n({\rm H_2})$ &	$2.13 \times 10^{-4}$&$2.50 \times 10^{-4}$&$2.37 \times 10^{-4}$&$2.35 \times 10^{-4}$&$1.39 \times 10^{-4}$\\ 
\hline
\multicolumn{6}{l}{$^a$ Haniff et al.\ (1995); estimates of $R_*$ were scaled in proportion to the assumed source distance} \\
\multicolumn{6}{l}{$^b$ Bergman \& Humphreys (2020); estimates of $R_*$ were scaled in proportion to the assumed source distance} \\
\multicolumn{6}{l}{$^c$ Matsuura et al.\ (2014); estimates of $R_*$ were scaled in proportion to the assumed source distance} \\
\multicolumn{6}{l}{$^d$ Khouri et al.\ (2014); estimates of $R_*$ were scaled in proportion to the assumed source distance} \\
\multicolumn{6}{l}{$^e$ Assumed the same as for W Hya} \\
\multicolumn{6}{l}{$^f$ $T = T_*(R/R_*)^{-0.15}$ for $R < 5 \times 10^{14}$~cm and $T \propto R^{-0.6}$ at larger $R$ (M14) } \\
\multicolumn{6}{l}{$^g$ Assumed equal to the sound speed at temperature $T_*$} \\
\multicolumn{6}{l}{$^h$ Standard value proposed by Maercker et al.\ 2016} \\
\multicolumn{6}{l}{$^i$ See Table 1 for references} \\
\multicolumn{6}{l}{$^j$ Density profile implies a constant $\dot M$, except for VY CMa.}\\ 
\multicolumn{6}{l}{\phantom{$^j$}  For VY CMa, the density profile (M14) implies a variable $\dot M$, which is matched to the value in Table 1}\\
\multicolumn{6}{l}{\phantom{$^j$} when $v =v_\infty$ }\\
\enddata
\end{deluxetable}

\end{document}